\documentclass[twocolumn]{aastex61}

\pdfoutput=1 
\usepackage{amsmath,amstext}
\usepackage[T1]{fontenc}
\usepackage{apjfonts} 
\usepackage[figure,figure*]{hypcap}

\usepackage{CJKutf8}
\AtBeginDvi{\input{zhwinfonts}} 
\newcommand{\Chi}[2]{%
  \csname CJK*\endcsname{UTF8}{zhsong}%
    \CJKchar{#1}{#2}%
  \csname endCJK*\endcsname
}
\DeclareUnicodeCharacter{9673}{\Chi{"96}{"73}}
\DeclareUnicodeCharacter{82F1}{\Chi{"82}{"F1}}
\DeclareUnicodeCharacter{540C}{\Chi{"54}{"0C}}
\DeclareUnicodeCharacter{6797}{\Chi{"67}{"97}}
\DeclareUnicodeCharacter{7701}{\Chi{"77}{"01}}
\DeclareUnicodeCharacter{6587}{\Chi{"65}{"87}}

\newcommand{\gsim}{\lower.7ex\hbox{$\;\stackrel{\textstyle>}{\sim}\;$}}
\newcommand{\lsim}{\lower.7ex\hbox{$\;\stackrel{\textstyle<}{\sim}\;$}}

\newcommand{\HELIOLINC}{{\sc HelioLinC}}

\shorttitle{Down the Back of the Couch}
\shortauthors{Holman et al.}

\submitjournal{ApJ/AJ May 2018}

\begin{document}

\title{Finding Asteroids Down the Back of the Couch:\\A Novel Approach to the Minor Planet Linking Problem}


\author[0000-0002-1139-4880]{Matthew~J.~Holman}
\affiliation{Harvard-Smithsonian Center for Astrophysics, 60 Garden St., MS 51, Cambridge, MA 02138, USA}
\affiliation{School of Engineering and Applied Sciences, Harvard University, 29 Oxford St., Cambridge, MA 02138, USA}
\correspondingauthor{Matthew~J.~Holman}
\email{mholman@cfa.harvard.edu}

\author[0000-0001-5133-6303]{Matthew~J.~Payne} 
\affiliation{Harvard-Smithsonian Center for Astrophysics, 60 Garden St., MS 51, Cambridge, MA 02138, USA}

\author{Paul Blankley}
\affiliation{School of Engineering and Applied Sciences, Harvard University, 29 Oxford St., Cambridge, MA 02138, USA}

\author{Ryan Janssen} 
\affiliation{School of Engineering and Applied Sciences, Harvard University, 29 Oxford St., Cambridge, MA 02138, USA}

\author[0000-0003-0645-6626]{Scott Kuindersma} 
\affiliation{School of Engineering and Applied Sciences, Harvard University, 29 Oxford St., Cambridge, MA 02138, USA}


\begin{abstract}
We present a novel approach to the minor planet linking problem.
Our heliocentric transformation-and-propagation algorithm clusters tracklets at common epochs, allowing for the efficient identification of tracklets that represent the same minor planet.
This algorithm scales as $O(N log N)$, with the number of tracklets $N$, a significant advance over standard methods, which scale as $O(N^3)$.  This overcomes one of the primary computational bottlenecks faced by current and future asteroid surveys.  We apply our algorithm to the Minor Planet Center's Isolated Tracklet File, establishing orbits for $\sim41,000$ new minor planets.
\end{abstract}

\keywords{%
}

\section{Introduction}
\label{SECN:INTRO}
A number of ongoing wide-field surveys, such as Pan-STARRS~\citep{Denneau.2013}, the Catalina Sky Survey~\citep{Christensen.2016}, OSSOS~\citep{Bannister.2016}, NEOWISE~\citep{Mainzer.2011}, and ZTF~\citep{Kulkarni.2016}, as well as planned surveys such as LSST~\citep{Jones.2017} and NEOCam~\citep{Mainzer.2017}, are designed to address a range of goals from constraining models of planet formation, through finding evidence of additional planets in our solar system \citep{Trujillo.2014, Sheppard.2016, Gerdes.2017}, to fulfilling the US Congressional mandate to discover 90\% of the potential hazardous asteroids with diameters exceeding 140m\footnote{National Aeronautics and Space Administration Authorization Act of 2005 (Public Law 109-155), January 4, 2005, Section 321, George E. Brown, Jr. Near-Earth Object Survey Act}.  

The typical survey strategy, as it relates to minor planets, is based on identifying `tracklets'.  
A tracklet is a sequence of two or more astrometic detections that are taken over a time span that is short enough that it is likely that the detections correspond to the same moving object, and long enough to allow solar system objects to be distinguished from stationary background sources.
A primary goal is to obtain a sufficient number tracklets for each object, distributed over a long enough time span, such that the resulting orbit is accurate enough to readily identify matching observations of the object in the past or future.  It is typically necessary to observe tracklets on three different nights to reliably establish an orbit for a main belt asteroid~\citep{Kubica.2007a,Denneau.2013,Jones.2017}. 

Current surveys observe wide areas of the sky and detect such large numbers of minor planets that it is not feasible to obtain follow up observations of each of the detected objects.  Some tracklets coincide with the predicted positions of already known minor planets with well determined orbits and thus do not require additional observations. Others, such as those that have locations and rates of motion that are consistent with being NEOs, are prioritized for immediate additional observation.  The rest must be linked with other tracklets from the same or other surveys, or they will be lost.  By design, most objects are naturally re-observed in the course of these surveys.  However, the tracklets corresponding to each object must still be identified before the orbits of those objects can be determined.   This is known as the `linking problem.'
 
The linking problem is challenging for several reasons.  First, asteroids are dense on the sky ($\sim~400\,deg^{-2}$ near the ecliptic at magnitude $r~\sim 24.5$).  Surveys can also produce significant numbers of false detections~\citep{Denneau.2013, Jones.2017}, which are incorporated into false tracklets that also contribute to the sky density.  The number of tracklets controls the amount of computation required. Thus, a higher sky density of tracklets results in a higher computational burden.  Second, the nearly power law distribution of minor planet sizes ensures that nearly all tracklets are  near the detection limits and thus cannot be easily distinguished by differences in apparent brightness.  Third, many tracklets have similar sky plane velocities, which also makes them difficult to distinguish.  Fourth, the apparent motion of minor planets is nonlinear over the span of months when observed from the Earth, therefore a tracklet may not point in the direction toward or away from its predessors and successors.

 The brute force solution to the linking problem would be to fit an orbit to every pair of tracklets.  For those pairs that yield a valid orbit, the remaining tracklets can be checked a third time for additional matches.  Given millions of tracklets and the computational cost of an individual orbit fit, the brute force approach is currently computationally intractable.  

The best available solution to the linking problem, the Pan-STARRS Moving Object Processing System (MOPS), is sophisticated but also complex~\citep{Kubica.2007a, Denneau.2013}.  After first identifying tracklets, MOPS projects each tracklet forward and backward in time, using expressions for RA and Dec that are quadratic in time, with predefined ranges of coefficients.   MOPS then uses KD-trees to efficiently identify other tracklets near those predicted locations.  Based on quadratic fits to pairs of tracklets, MOPS searches for matching third tracklets.  The resulting candidate groups of three tracklets are then tested with orbit fitting, which dependably verifies if the tracklets correspond to the same object, with low false positive and false negative rates.  MOPS achieves a high level of completeness in simulations~\citep{Denneau.2013, Veres.2017a,Veres.2017b, Jones.2017}.

Despite these advances, the MOPS approach is still a variation of the brute force method, bringing groups of three tracklets together to be tested with orbit fitting.  Predicting the location of plausibly matching tracklets and using a KD-tree to efficiently locate those tracklets significantly improve the overall efficiency, but the number of orbit fits that must be carried out, which is the most computationally intensive step, still scales as $\mathcal{O}(N_t^3)$, where $N_t$ is the number of tracklets~(see eq.~A22 of \citealt{Jones.2017}).  
LSST is planning to dedicate $\sim1000$~CPUs to identifying and linking asteroid tracklets with MOPS~\citep{Jones.2017}.
Although this is a small fraction of the computational resources available to LSST, it illustrates the scale of the linking problem using currently available solutions.

Fortunately, we can exploit a useful characteristic of short-arc asteroid orbits to develop a more efficient method.
The parameters of such orbits can be neatly separated into those that are well determined and those that are poorly known. 
The orbit of any minor planet can be described with six parameters (three position components and three velocity components) at a reference time.  
Observations of a single tracklet provide precise estimates of four of these: the sky plane location and two angular rates of motion.  
However, the topocentric distance $\rho$ and radial velocity $\dot\rho$ are not directly observed with astrometry and are poorly known, initially. The fundamental challenge in the linking problem, and with orbit fitting in general, is to infer the distance to the object at the times of the observations. 

 For asteroids, $\rho$ can vary widely and rapidly.  For a near-Earth object (NEO) making a close approach to the Earth, $\rho$ can vary by orders of magnitude in the course of days.  {\it A priori} estimates of $\rho$ are not well defined, unless diurnal parallax is evident.  

On the other hand, the heliocentric distance $r$ is slowly varying and has a well-posed prior distribution.
If one were able to observe from the Sun, the minor planets would appear to trace great circles on the sky, locally following straight lines in heliocentric angular coordinates.  Moreover, the angular velocity of the motion along this great circle would be a simple function of true anomaly, reaching its minimum at apocenter and maximum at pericenter.  

A productive approach is to simply assert a set of values for the unknown heliocentric distance.  For each of these assumed distances, one can transform the observations to a heliocentric frame and then search for great circle motion.  The observations of objects that are actually near the asserted distance will line up.  We developed this approach, which we call `heliocentric linking', and have successfully applied it to searches for distance solar system objects in time-sparse Pan-STARRS data~\citep{Chen.2016,Lin.2016, Holman.2017}.  Such a method was recently elaborated upon and applied to a search of WISE data for distant objects~\citep{Perdelwitz.2018}.  
We note that the heliocentric linking approach, whether applied to single detection or tracklets, still scales as $\mathcal{O}(N^m)$, where $N$ is the number of detections or tracklets to be analyzed and $m$ is the number required to make a confident discovery ($m\sim3-5$).

However, if one knew the missing information for each tracklet, namely both $r$ and $\dot r$ or equivalently $\rho$ and $\dot \rho$, one would have a full specification of the dynamical state and could integrate the tracklet trajectories to a common time.  Those tracklets that correspond to the same object would coincide in position and velocity, to within the observational uncertainties, because they have the same underlying orbit.   One could then search for clusters to identify which tracklets correspond to the same object.

In this paper, we combine these two ideas, heliocentric linking and clustering of tracklets, into a novel and efficient solution to the linking problem.
We refer to this method as \HELIOLINC.
In \S\ref{SECN:HELIO}, we describe the \HELIOLINC\ method. %
In \S\ref{SECN:ALGO}, we describe our algorithm for identifying clusters of tracklets (within sets of transformed tracklets).
In \S\ref{SECN:DEMO}, we demonstrate the training and performance of the \HELIOLINC\ algorithm on previously identified tracklets in the Minor Planet Center's database of Unnumbered objects.
In \S\ref{SECN:ITF}, we employ our algorithm to identify new objects within the Minor Planet Center's ``Isolated Tracklet File'',
and then in \S\ref{SECN:DISCUSS}, we discuss the implications of our results for ongoing and future surveys.

\section{Transformations}
\label{SECN:HELIO}

We follow the formalism and notation of \citet{Bernstein.2000}, with key changes that we will highlight.  We consider a minor planet orbiting the Sun.  Its position in inertial space at time $t$ is given by ${\bf x}(t)$.  That of the observatory, ${\bf x_E}(t)$, is known precisely.  The coordinate system has the z-axis pointed outward toward a location on the sky and the x-y plane is perpendicular to that, coinciding with the local sky plane.  (By convention, the x-axis is parallel to the direction of increasing ecliptic longitude, and the y-axis completes a right-handed system.)   
\citet{Bernstein.2000} generally adopt a coordinate system that is oriented with the z-axis in the direction of the first observation, and the origin is located at the observatory at the time of the first observation.  
Instead, we divide the sky into regions, using the HEALPix tessellation~\citep{Gorski.2005}, take the center of each as the reference direction for a local sky region, and place the origin at the Sun (or barycenter).
\citet{Bernstein.2000} take the reference time, $t=0$, to be the time of the first observation in a tracklet or set of tracklets.  Instead, we adopt a common reference time for all tracklets that we will attempt to link.  For example, we might take as the reference time the date of new moon for the month being considered.  
Choosing a common reference time and coordinate system for a set of tracklets that are to be linked is a key part of our method. As will be seen, this approach allows us to efficiently determine which tracklets might correspond to the same minor planet.

As stated in \citet{Bernstein.2000}, the observed angular coordinates of an asteroid in the local tangent plane are given by
\begin{equation}
\begin{array}{ccc}
\theta_x(t) & = & { {x(t^\prime) - x_E(t)} \over {z(t^\prime) -
z_E(t)}} \\
\theta_y(t) & = & { {y(t^\prime) - y_E(t)} \over {z(t^\prime) -
z_E(t)}, }
\end{array}
\label{exact}
\end{equation}
where $t^\prime=t-\Delta t$, and $\Delta t$ is the light travel time from the object to the observer.

The trajectory of the target body can be separated into a linear portion and a gravitational perturbation: 
\begin{equation}
{\bf x}(t) = {\bf x}_0 + {\bf \dot x}_0 t + {\bf g}(t). 
\end{equation}
The gravitational perturbation ${\bf g}(t)$  is given by 
\begin{equation}
{\bf g}(t=0)  = 0     \>\ , \>\ 
\dot {\bf g}(t=0) = 0 \>\ , \>\
\ddot {\bf g}(t)  \approx  -GM_\odot \frac{{\bf x}(t)}{|{\bf x}(t)|^3 },
\label{eqn:g}
\end{equation}
where $GM_\odot$ is the gravitational constant of the Sun.  It is worth noting that ${\bf g}(t)$ is small for $t\ll~T_{orb}$, where $T_{orb}$ is the orbital period of the object.
We have ignored the perturbations of the planets and massive asteroids in Equation \ref{eqn:g}, as they are even smaller than the perturbation from the Sun.  However, they can easily be included.

\citet{Bernstein.2000} introduce the following helpful parameterization, based on the components of the inertial position and velocity of the target at the reference time:
\begin{equation}
\label{definitions}
\begin{array}{lllll}  
\alpha \equiv {x_0}/{z_0} &,& \beta \equiv {y_0}/{z_0} &,& 
	\gamma \equiv 1/{z_0} \\ 
\dot\alpha \equiv {\dot x_0}/{z_0}&,& \dot\beta \equiv {\dot y_0}/{z_0} &,& 
	\dot\gamma \equiv {\dot z_0}/{z_0}.
\end{array}
\end{equation}
In this system, $\alpha$ and $\beta$ are the components of the angular position of the object at the reference time,  $\dot \alpha$ and $\dot \beta$ are angular rates of motion in the inertial coordinate system, $\gamma$ is a measure of distance to the object, and $\dot \gamma$ is a scaled radial velocity.   
It is worth noting that the dotted parameters are the velocity components scaled by $z_0$, rather than time derivatives.
In terms of these parameters,  the observations $\theta_x(t)$ and $\theta_y$ are:
\begin{equation}
    \begin{split}
        \theta_x & =  { { \alpha + \dot\alpha t^\prime + \gamma g_x(t^\prime) - \gamma x_E(t) }
        \over { 1 + \dot\gamma t^\prime + \gamma g_z(t^\prime) - \gamma z_E(t) } }
        \\
        \theta_y & = { { \beta + \dot\beta t^\prime + \gamma g_y(t^\prime) - \gamma y_E(t) }
        \over { 1 + \dot\gamma t^\prime + \gamma g_z(t^\prime) - \gamma z_E(t) } },
    \end{split}
\label{posexpression}
\end{equation}
where $t^\prime \approx t - \frac{1}{c \gamma}$ is the light-time corrected time of the observation.

The observations of a tracklet constrain four of the six quantities needed to specify an orbit: two angular positions and two angular rates. \citet{Bernstein.2000} note the total degeneracy between $\dot\alpha$ and $\gamma \dot x_E$ for observations near opposition.  Nearer targets with small transverse velocity have the same apparent angular rate of motion as more distant targets with large transverse velocities.  Although this degeneracy limits the quality of orbit fits if observations are restricted to short arcs near opposition, it is advantageous for the linking problem.  
 This degeneracy reduces the dependence on $\gamma$: errors in $\gamma$ can be absorbed by changes in $\dot\alpha$.
Furthermore, the expressions for $\theta_x$ and $\theta_y$ are nearly linear in the parameters, a feature designed and highlighted by \citet{Bernstein.2000}.

In our earlier work with heliocentric linking we searched a set of heliocentric distances~\citep{Lin.2016,Chen.2016,Holman.2017}.  In the present work, we assume values for {\it both} the distance and its rate of change through $\gamma$ and $\dot\gamma$.  We can rearrange equations~\ref{posexpression} to yield simple expressions for the linear motion of the object:
\begin{equation}
    \begin{split}
        { { \alpha + \dot\alpha t^\prime }} = 
                    &\, \theta_x \left[ 1 + \dot\gamma t^\prime + \gamma g_z(t^\prime) - \gamma z_E(t)\right] 
                    \\ 
                    & -\gamma g_x(t^\prime) + \gamma x_E(t)
        \\
        { { \beta + \dot\beta t^\prime }}  = 
                    &\, \theta_y \left[ 1 + \dot\gamma t^\prime + \gamma g_z(t^\prime) - \gamma z_E(t)\right] 
                    \\ 
                    & -\gamma g_y(t^\prime) + \gamma y_E(t),
    \end{split}
\label{rearrange}
\end{equation}
where $\theta_x$ and $\theta_y$ are observed quantities, and the observatory position ($x_E$, $y_E$, $z_E$) is known precisely.  We note that in Equation \ref{rearrange} the transverse components of the gravitational perturbation, $g_x(t^\prime)$ and $g_y(t^\prime)$, are much smaller than $g_z(t^\prime)$.   Furthermore, the two equations are independent of each other, if $\gamma$ and $\dot\gamma$ are assumed.   The factor in brackets is the same in both expressions.  These properties simplify the solution.

\begin{figure}[htbp]
\centering
  \includegraphics[trim = 0mm 0mm 0mm 0mm, clip, angle=0, width=0.65\columnwidth]{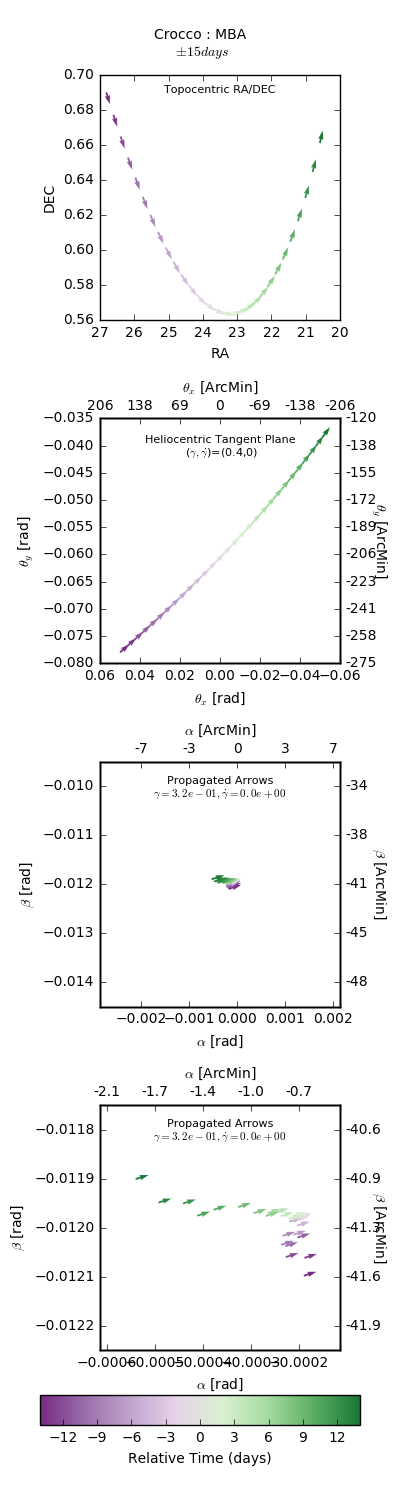}
  \caption{%
    Illustration of transformations on the MBA Crocco.
    {\bf Top:} Topocentric Equatorial RA,Dec observations: note retrograde motion. 
    {\bf Middle:} Transformed $\theta_x$ and $\theta_y$ coordinates as per Eqn. \ref{posexpression}, illustrating prograde motion. 
    {\bf Bottom:} Effect of propagating ``arrows'' to a common epoch (Eqn. \ref{rearrange}), illustrating the extremely tight resultant clustering (common scales deliberately selected for middle and bottom plots). 
    %
    }
    \label{fig:TRANSF}
\end{figure}

As described below, we will carry out least squares fits of equations~\ref{rearrange} for each tracklet to obtain the parameters $\alpha$, $\dot\alpha$, $\beta$, and $\dot\beta$.  These parameters represent components of the motion in inertial space in a common reference frame at a common reference time. Thus, they can be compared to determine which tracklets correspond to the same underlying object.   We refer to a set of these four parameters as an `arrow' to distinguish it from a `tracklet', which refers to the original set of observations in sky coordinates~\citep{Kubica.2007a}, and an `attributable', which refers to the parameters resulting from a linear fit of a tracklet in sky coordinates at the time of the tracklet~\citep{Milani.1999}.

In Figure \ref{fig:TRANSF} we demonstrate the effects of the  transformation and fitting outlined in Equations~\ref{posexpression} and \ref{rearrange} on tracklets for the known Main Belt Asteroid (10606) Crocco. 

The top panel of Figure~\ref{fig:TRANSF} shows a time series of tracklets for (10606) Crocco 
in topocentric coordinates (RA/Dec) near opposition.  The sequence of tracklets shows apparent retrograde motion, as well as curvature (which is exaggerated by the vertical scale of the panel).  The next lower panel shows those same tracklets projected onto a local tangent plane.  The subsequent panel shows the tracklets after they have been transformed to heliocentric coordinates with an assumed value of $\gamma=0.4$ ($r=2.5$~AU).  In heliocentric coordinates, the tracklets appear to line up, following a great circle.  The bottom panel shows the arrows $(\alpha, \dot\alpha, \beta, \dot\beta)$ that result from fitting the tracklets, assuming $\gamma=0.4$ and $\dot\gamma=0$.
Choosing the reference frame this way means that at the reference time the values of $\theta_x$ and $\theta_y$ are zero, and they diverge from this in an approximately linear manner for observations at times either side of this reference time.
By propagating the arrows back to the reference epoch, we see in the bottom panel of Figure \ref{fig:TRANSF} that the tracklets cluster within a small radius.

\subsection{Gravitational Perturbation}
Before continuing, we now examine the gravitational perturbation $\bf{g}(t)$ in more detail.  The position and velocity vectors of the target can be represented as 
\begin{eqnarray}
{\bf x}(t) &=& \emph{\tt f$_{\tt G}$}(t)\, {\bf x}_0 + \emph{\tt g$_{\tt G}$}(t) \, {\bf \dot x}_0\\
{\bf \dot x}(t) &=& 
\dot{\emph{\tt f}}_{\tt G}(t)\, {\bf x}_0 + 
\dot{\emph{\tt g}}_{\tt G}(t) \, {\bf \dot x}_0,\nonumber
\end{eqnarray}

where $\emph{\tt f$_{\tt G}$}(t)$ and $\emph{\tt g$_{\tt G}$}(t)$ are the Gauss `f' and `g' functions, and $\dot{\emph{\tt f}}_{\tt G}(t)$ and 
$\dot{\emph{\tt g}}_{\tt G}(t)$ are their time derivatives~\citep{Danby.1992}.  
The $\emph{\tt f$_{\tt G}$}(t)$ and $\emph{\tt g$_{\tt G}$}(t)$ functions describe Keplerian motion about the Sun in the plane defined by ${\bf x}_0$ and ${\bf \dot x}_0$ (The perturbations from the planets are far smaller.) 
The gravitational perturbation of equation~\ref{eqn:g} is then
\begin{eqnarray}
{\bf g}(t)  &=& \emph{\tt f$_{\tt G}$}(t)\, {\bf x}_0 + \emph{\tt g$_{\tt G}$}(t) \, {\bf \dot{x}}_0 - [{\bf x}_0 - {\bf\dot{x}}_0 t ]
\\
            &=& [ \emph{\tt f$_{\tt G}$}(t)-1]\, {\bf x}_0 + [\emph{\tt g$_{\tt G}$}(t)-t] \, {\bf\dot{x}}_0.\nonumber
\end{eqnarray}
The time-dependent coefficients can be approximated with well known series:
\begin{eqnarray}
\emph{\tt f$_{\tt G}$}(t)-1 &=& -\frac{1}{2} \sigma t^2 + \frac{1}{2}\sigma \tau t^3 + ...\nonumber\\
\emph{\tt g$_{\tt G}$}(t)-t &=& -\frac{1}{6} \sigma t^3 + ...\nonumber
\end{eqnarray}
with $\sigma = GM_\odot/r_0^3$ and $\tau = \dot{r}_0/r_0$~\citep{Danby.1992}.  In our basis, $\sigma = GM_\odot \gamma^3$ and $\tau\approx\dot\gamma$, to a high degree of accuracy.  Thus,
\begin{eqnarray}
{\bf g}(t) &\approx& (-\frac{1}{2} \sigma t^2 + \frac{1}{2}\sigma\tau t^3 ) {\bf x}_0 + (-\frac{1}{6}\sigma t^3 ) {\bf \dot x}_0,
\end{eqnarray}
where the $t^3$ terms represent the gravitational jerk. 

In terms of our parameters, the expressions relevant to equations~\ref{posexpression} and \ref{rearrange} are:
\begin{eqnarray}
\gamma g_x(t) &\approx& -\frac{1}{2} \sigma t^2 \alpha - \frac{1}{6}\sigma\tau t^3 (\dot\alpha - 3 \alpha),\\
\gamma g_y(t) &\approx& -\frac{1}{2} \sigma t^2 \beta - \frac{1}{6}\sigma\tau t^3 (\dot\beta - 3 \beta),\nonumber\\
\gamma g_z(t) &\approx& -\frac{1}{2} \sigma t^2 - \frac{1}{6}\sigma\tau t^3 (\dot\gamma - 3).\nonumber
\end{eqnarray}
There are a few options for the gravitational perturbation, in order of increasing accuracy and computational cost:
\begin{itemize}
    \item Ignore it.  The leading order terms are small and $O(t^2)$, so entirely neglecting the gravitational perturbation is reasonable if the time span is short enough.
    \item Neglect all terms except for $g_z(t) \approx -\frac{1}{2} \sigma t^2$. The other terms are even smaller or higher order in $t$.  This approach has the advantage that it only depends upon $\alpha$, $\beta$, and $\gamma$, but not the other parameters.   It also requires negligible additional computation.
    \item Include all of the terms listed above.  This necessitates iteration in the fitting of arrows, because $\alpha$, $\dot\alpha$, $\beta$, and $\dot\beta$ are needed to evaluate the perturbation in this approximation.
    \item Exactly solve the Kepler step, rather than using series expansions for $\emph{\tt f$_{\tt G}$}(t)$ and $\emph{\tt g$_{\tt G}$}(t)$.  This necessitates both an iteration to solve Kepler's equation, as well as an overall iteration to fit for the arrows.
    \item Include all the gravitational perturbations from the Sun and planets.  This necessitates an n-body integration for each iteration in the fitting for the arrows.
\end{itemize}

We adopt the second option for most of our calculations.  Examining the other options is left for future work.

\section{Clustering Algorithm}
\label{SECN:ALGO}

\subsection{Description of Algorithm}
\label{SECN:ALGO:DESC}


In this section we describe the details of each stage of the algorithm.  These include preliminary calculations, a series of transformations, finding clusters, and verifying those clusters.


\subsubsection{\sc Preliminary Calculations}
\label{SECN:ALGO:DESC:PRELIM}
Most surveys concentrate on observing the regions of sky near solar opposition during dark time.   They typically re-observe the same regions of sky a few times during a single lunation.
By design, often there are enough tracklets observed of individual objects to support linking using just the observations from a single lunation.  
This matches the proposed observing strategy for LSST, as well its goals for linking tracklets over 12-15~day intervals.
Thus, we divide each data set into  $\pm 15$~day windows centered on the dates of new moon, which produces units of data that can be processed separately.    

For each time window, we further spatially divide the tracklets into separate heliocentric sky regions. 
The sky regions are chosen to be sufficiently large that a minor planet cannot traverse an entire region in the time span being considered.  
As will be shown below, the results are relatively insensitive to the choice of distance.  We find that only a few values of $\gamma$ (the inverse of the heliocentric distance), that match the rough distance classes of solar system bodies, are needed to achieve good results.

To determine the heliocentric position vector ${\bf r}$ of the first detection in each tracklet, we assume the value of $\gamma$ of the distance class, such as $\gamma=0.4$ ($r=2.5$~AU), and transform the topocentric position vector to its heliocentric counterpart.  We refer the reader to Appendix \ref{APP:HELIO} for details of the heliocentric transformation. 

Given the heliocentric position vector ${\bf r}$, we determine a spatial index for the tracklet using the HEALPix sky tessellation~\citep{Gorski.2005}.
The details of the tessellation are not important for our application.  For our purposes, HEALPix is simply a convenient means of spatially organizing the data into sky regions with accompanying central reference directions.  Other tessellations such as Hierarchical Triangular Mesh (HTM)~\citep{Szalay.2007} are suitable.

At this point, for each distance class, we have divided the tracklets into time windows and sky regions. 

\subsubsection{ \sc Transformations }
\label{SECN:ALGO:DESC:TRANS}

The tracklets in each combination of time window and sky region form a data set that can be processed independently.  The reference time is the center of the time window for the data set, and the reference direction is direction to the center of the sky region.

For each of these data sets, we iterate over a set of $(\gamma,\dot{\gamma})$ parameters to carry out the transformations described in Section \ref{SECN:HELIO}.  For each $(\gamma, \dot\gamma)$ pair, we transform and least squares fit each tracklet in the data set according to equations~\ref{rearrange}.  The result is an `arrow', the set of $(\alpha$, $\dot\alpha$, $\beta$, $\dot\beta)$ parameters, for each tracklet. 
For the $(\gamma, \dot\gamma)$ combination that corresponds to the actual orbit of an object, the transformed arrows for that object will coincide in the space of $(\alpha$, $\dot\alpha$, $\beta$, $\dot\beta)$.  

\begin{figure}[htbp]
\centering
  \includegraphics[trim = 0mm 0mm 0mm 0mm, clip, angle=0, width=0.95\columnwidth]{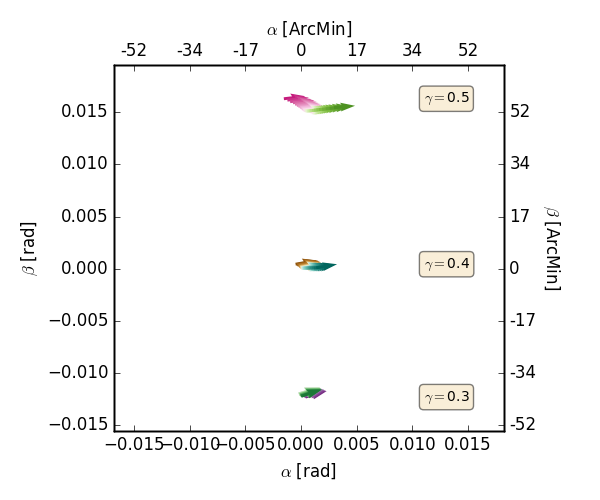}
  \includegraphics[trim = 0mm 0mm 0mm 0mm, clip, angle=0, width=0.95\columnwidth]{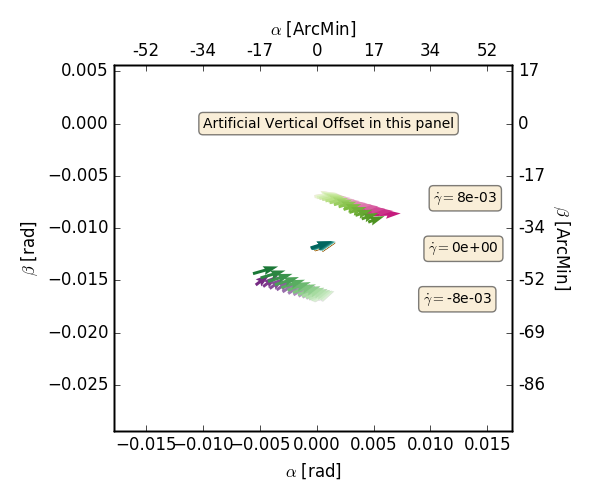}
  \caption{%
    Illustration of varying gamma (top) and gamma-dot (bottom) to show that the tightness of the clusters is relatively insensitive to gamma but quite sensitive to gamma-dot.
    N.B. The top and bottom sets of $\dot\gamma$ in the bottom plot have each received relative offsets of $5\times10^{-3}$ radians in the y-direction for clarity. 
    }
    \label{fig:TRANSF_v2}
\end{figure}

How close to the actual values must the assumed values for $\gamma$ and $\dot\gamma$ be for a cluster of arrows to be apparent?  The spacing in the assumed parameters should be fine enough to ensure that the uncertainties of the arrows are dominated by the observational uncertainties, rather than by the granularity of the assumed parameters.  Figure~\ref{fig:TRANSF_v2} illustrates how the distribution of arrows depends up $\gamma$ and $\dot \gamma$.   In the top panel, the arrows are shown for several assumed values of $\gamma$, while keeping $\dot \gamma$ fixed at the correct value ($\gamma\sim 0.3)$.  The cluster spreads out as the assumed values $\gamma$ depart farther from the correct value.  In the bottom panel, the arrows are show for several assumed values of $\dot \gamma$, keeping $\gamma$ fixed at the correct value.  The panels illustrate that relatively tight clusters can be formed without extremely fine steps in $\gamma$ and $\dot \gamma$.  We explore this empirically  in Section~\ref{SECN:DEMO}.

\subsubsection{ \sc Make Clusters}
\label{SECN:ALGO:DESC:CLUST}

We link tracklets by identifying clusters of their corresponding arrows.   There is a wide variety of clustering algorithms; a full exploration of clustering methods is beyond the scope of this work.  For this paper, we use a simple KD-tree approach. KD-trees are especially suited to nearest-neighbour detection and are fast for low-dimensional data~\citep{Kubica.2007a}. 
We use a dual-tree algorithm to efficiently find the neighbors within a given cluster radius $d$ of every arrow in the tree~\citep{Curtin.2017}.

We populate a four-dimensional KD-tree with the arrow parameters ($\alpha$, $\beta$, $\dot\alpha$, $\dot\beta$) determined for the assumed parameters $\gamma$ and $\dot\gamma$.
We use the following metric for the distance between two arrows (distinguished by unprimed and primed parameters):
\begin{eqnarray}
    d^2 &=&
    (\alpha-\alpha^\prime)^2 + (\beta-\beta^\prime)^2 + \\\nonumber
    && dt^2 \left[ (\dot\alpha-\dot\alpha^\prime)^2 + (\dot\beta-\dot\beta^\prime)^2 \right],
    \label{eqn:d}
\end{eqnarray}
where $dt$ is a constant factor, in units of time, that relates the angular velocities and the angular positions.  
The scale factor $dt$ and the cluster radius $d$ are hyper-parameters that we train in Section~\ref{SECN:DEMO}.

Arrows that are separated by less than a specified distance will be both close in angular space and moving parallel to each other.
Arrows that correspond to the same object, for which the grid parameters $\gamma$ and $\dot\gamma$ are close, will form a tight cluster.
We define an acceptable, preliminary cluster as having three or more tracklets within the specified radius.  This is consistent with the objective of other linking methods.

\subsubsection{ \sc Verify Clusters}
\label{SECN:ALGO:DESC:VERIFY}
The identified clusters need to be examined to verify that the constituent tracklets are all consistent with corresponding to a single object in heliocentric orbit. To verify that this is the case, we undertake a number of tests designed to exclude obvious ``contaminant'' tracklets, performing the simpler tests first.
We 
(i) eliminate tracklets with duplicate times, 
(ii) eliminate inconsistent time/space orderings, 
and 
(iii) perform full orbit fits. 
Orbiting fitting is relatively computationally expensive, however it is sufficiently efficient to perform over the list of tracklets produced by the KD-tree clustering once they have been cleaned-up.  
Therefore, each matched cluster that passes the above tests has an orbit fit performed to verify it indeed matches a realistic heliocentric orbit.

Further details on the verification methods employed are provided in Section \ref{s:validation}.

\subsection{Algorithmic Scaling}
\label{s:algo:scaling}

It is important to note that the algorithm 
described in Section \ref{SECN:ALGO:DESC}, 
contains \emph{no} nested loops over tracklets.
This is of crucial importance, as it means that the algorithmic compute time is essentially linear in the number of tracklets, $\mathcal{O}(N_t)$.
We note that clustering of tracklets scales as $\mathcal{O}(N_c\log N_t)$, where $N_c$ is the number of clusters. 
Therefore our algorithm will, at worst, scale as $\mathcal{O}(N_t\log N_t)$ in the limit that $N_c\rightarrow N_t$.
This is vitally important, providing hugely improved scaling over the $\mathcal{O}(N_t^3)$ scaling of algorithms such as  {\sc MOPS}.

Although it is natural to assume that $N_c<N_t$, where $N_c$ and $N_t$ are the number of clusters and input tracklets, respectively, it is possible for a tracket/arrow to belong to more than one cluster.   However, in almost all cases, and with proper clustering algorithms, $N_c\ll N_t$ .

It is worth noting that our algorithm has a large, but fixed, pre-factor.  
That is, for every tracklet we need to loop over a grid of $\gamma,\dot{\gamma}$ transformations.
But the size of this prefactor is small in comparison to the savings gained from reducing to an $\mathcal{O}(N_t\log N_t)$ scaling with number of tracklets.

It is also worth noting that all the calculations presented in this work were completed on single processor machines in less than a day.

\section{Demonstration of Algorithm:\\Labelled Data}
\label{SECN:DEMO}
We demonstrate our approach by linking tracklets from the Minor Planet Center's (MPC) data sets.  
In particular, we use the MPC's ``Unnumbered Observations'' file\footnote{\url{https://www.minorplanetcenter.net/iau/ECS/\\MPCAT-OBS/midmonth/UnnObs.txt.gz}}.  
Unnumbered minor planets typically have a large number of observations, spanning a few to many years.  
They have well determined orbits, but the quality of their orbits is not yet that of minor planets that have received numbered designations from the MPC.  

We extracted the most recent $10^7$ detections from the Unnumbered Observations file, at the time of processing, and selected from these only those objects for which there are at least 20 observations and at least three tracklets. 

We created the tracklets by grouping the observations with the same MPC provisional designation (in packed form, {\it i.e.} K15BH1W), the same observatory code ({\it i.e.} F51), and from the same 24-hour period (truncated MJD).  The tuple of provisional designation, observatory code, and truncated MJD uniquely identifies each tracklet.  This is how the MPC generally defines tracklets, although there are special cases. 

This process creates a sample of $\sim 1.4\times 10^5$ objects, composed of $\sim1.7\times10^6$ tracklets, containing a total of $6.4\times 10^6$ individual observations.  Because the identity of the tracklets is known from the provisional designation, we can use this sample as a labelled data-set that allows us to check the accuracy and completeness  of our clustering algorithm.

\subsection{Identifying Clusters in the Unnumbered Observation File: Fixed $\gamma = 0.4$}
\label{s:FixedGamma}
%
\begin{figure}[htbp]
\centering
    \includegraphics[trim = 0mm 35mm 0mm 30mm, clip, angle=0,     width=\columnwidth]{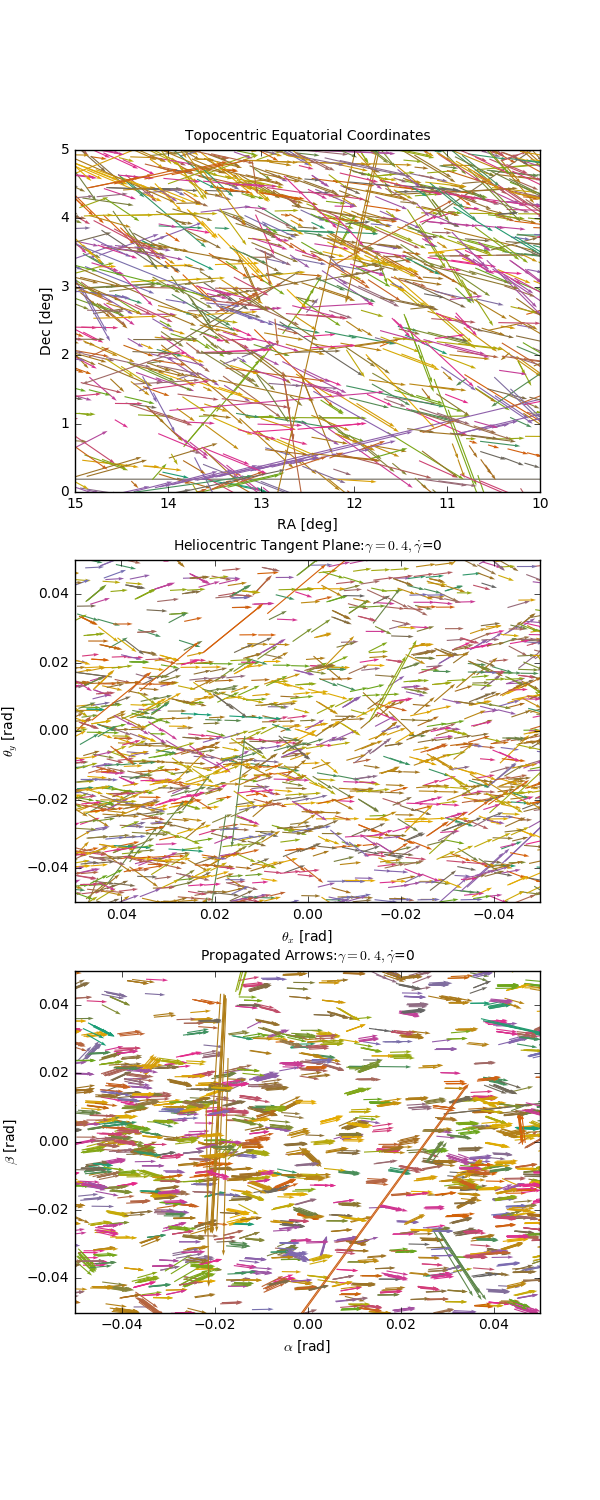}
    \caption{%
    {\bf Top:} Tracklets in a small region of the sky at opposition, near the ecliptic, plotted in equatorial coordinates. 
    {\bf Middle:} Tracklets transformed to heliocentric $\theta_x$ and $\theta_y$ coordinates as per Eqn. \ref{posexpression}, assuming $(\gamma,\dot\gamma)=(0.4,0)$.
    The direction and length of the arrows represent $\dot\theta_x$ and $\dot\theta_y$.
    {\bf Bottom:}
    Using Eqn. \ref{rearrange} we fit for ($\alpha,\beta,\dot\alpha,\dot\beta$).
    The tails of the plotted arrows indicate $\alpha$ and $\beta$, the direction and length represent $\dot\alpha$ and $\dot\beta$.  
    The arrow colors encode the identity of the objects: the propagated arrows in the bottom panel display a clear ``clustering'' by color, as the arrows from the same object become concentrated in the same region of parameter space.
    N.B. Fig. \ref{fig:focus} `zooms-in' on the bottom panel of this plot.    }
    \label{fig:UnnObs}
\end{figure}
%
In Figure \ref{fig:UnnObs} we illustrate our approach using a small sample of data from a region of sky at opposition, near the ecliptic. 
We use the known identities to plot tracklets from the same object with the same color. 
The top panel of Figure~\ref{fig:UnnObs} shows the tracklets in topocentric equatorial coordinates.  
This is how the tracklets are observed, as a collection of observations that define a sky plane location and direction of motion, {\it i.e. an attributable}~\citep{Milani.1999}. 
In these coordinates it is difficult to visually identify tracklets that correspond to the same object, although some groups are apparent.   

The middle panel shows the tracklets transformed to heliocentric tangent plane coordinates, assuming $\gamma=0.4$.  
In these coordinates, tracklets for objects with actual heliocentric distances that roughly correspond to $\gamma=0.4$ follow great circle motion on the sky, or straight-line motion in the tangent plane.  
Close inspection shows transformed tracklets that correspond to the same object. 

The bottom panel shows the corresponding {\it arrows} for those tracklets, assuming $\gamma=0.4$ and $\dot\gamma=0$.  
These essentially show the location direction of motion of the tracklets at the reference time.  Clusters of arrows are now readily apparent.  The arrows in these clusters correspond to the same object.
%
\begin{figure}[htbp]
\begin{minipage}[b]{\columnwidth}
\centering
    \includegraphics[trim = 0mm 35mm 0mm 35mm, clip, angle=0, width=0.8\columnwidth]{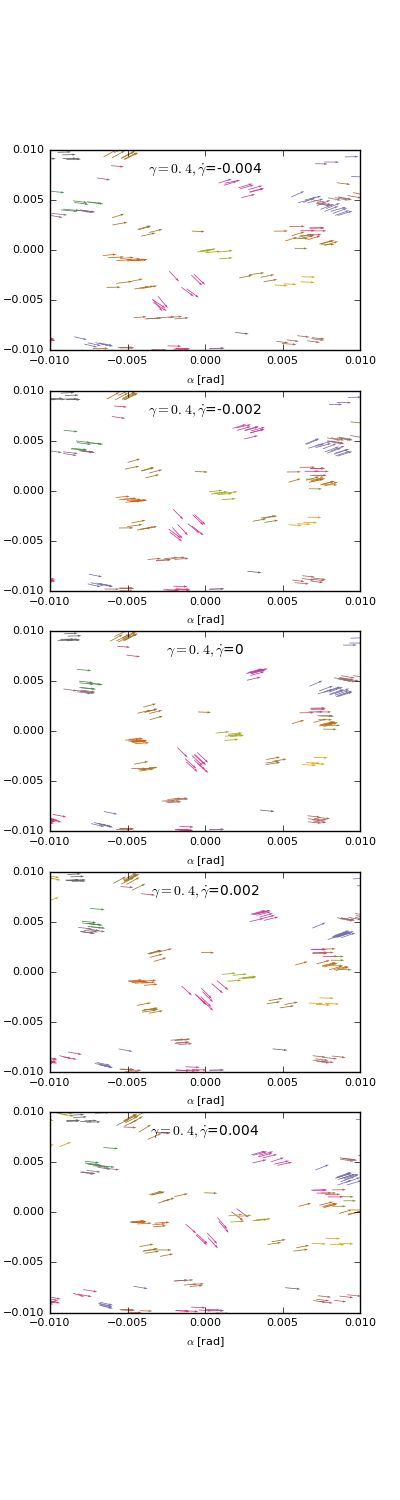}
    \caption{%
    Arrows created using different $\dot\gamma$.
    Arrows from the same known objects are plotted using the same color. %
    Scanning through different values of $\dot\gamma$ (top-to-bottom), we find that  clusters ``move in and out of focus'' as different radial velocities are asserted.
    N.B. the center panel of this plot is a `zoom-in' of the bottom panel in Figure \ref{fig:UnnObs}.
    }
    \label{fig:focus}
\end{minipage}
\end{figure}

In Figure \ref{fig:focus} we illustrate the manner in which different assumed values of $\dot\gamma$ affect the clustering of arrows.  The clusters become tightest at the value of $\dot\gamma$ closest to the true value of $\dot\gamma$ at the reference epoch. 

This illustrates our approach.  We examine a range of $\gamma$ and $\dot\gamma$ values.  For each data set, and for each assumed pair of $\gamma$, $\dot\gamma$ values, we proceed as outlined in Section \ref{SECN:ALGO:DESC:TRANS}, performing the transformation and least squares fit for each tracklet to determine its corresponding arrow parameters ($\alpha$, $\dot\alpha$, $\beta$, $\dot\beta$), as per Equation \ref{rearrange}.  We then search for clusters among those arrows.  Tight clusters of arrows correspond to the same object.

\subsection{Training Clustering Hyper-Parameters}
\label{SECN:HYPER1}   

The clustering algorithm depends upon two hyper-parameters:
\begin{itemize}
    \item $dt$: Velocity weighting of the cluster.  This controls the relative importance of the angular positions and angular velocities in the arrow distance metric.
    \item $d$: Clustering radius in four-dimensional space.
\end{itemize}

To optimize the parameters, we analyze the observations within $\pm15$~days of the center of five different, non-sequential lunations.
For each lunation, we undertake the preliminary calculations described in Section \ref{SECN:ALGO:DESC:PRELIM}, separating the data into equal area regions centered on each of the HEALPix centers ($n_{side}=8$).  The time of the center of the lunation and the unit vector to each HEALPix center establish the reference system for each data set.
Each of the regions includes its central HEALPix region and the adjacent eight regions.  This results in significant overlap between neighboring search regions.  This results in redundant calculations; optimizing the degree of overlap is left to future work.

As a first step, we pick a single value $\gamma=0.4$, which corresponds to the middle of the main asteroid belt ($r=2.5$~AU). 
We explore five  values of $\dot\gamma$ evenly spaced from $-4\times10^{-3}$ to $4\times10^{-3}$~rad/day.  
The extreme values of $\dot\gamma$ exceed the range for bound orbits for some parts of the main asteroid belt, thus allowing clustering of interstellar objects~\citep{Meech.2017}.

For a single value of $dt$ we examined the result of using a
Finer granularity in $\dot\gamma$ does not appear to be necessary in order to achieve a high linking efficiency for this particular data set.

We identify clusters in those parameters using the methods described in Section \ref{SECN:ALGO:DESC:CLUST}.
We define a cluster to consist of three or more arrows within the radius, $d$ (see Equation \ref{eqn:d}).  
We consider \emph{three} tracklets to be the minimum for successful clustering for this labelled data.  
As in the case of \emph{un}labelled data, clusters of three or more could be verified by explicit orbit-fitting.  
At this stage we are not concerned with the temporal distribution of the tracklets.  
It is possible that all three tracklets come from the same night.  
In such a case, the arc length would generally not be long enough to determine an orbit, but the tracklets would still support the veracity of the link.

\begin{figure}[htbp]
\centering
    \includegraphics[trim = 0mm 0mm 0mm 0mm, clip, angle=0, width=\columnwidth]{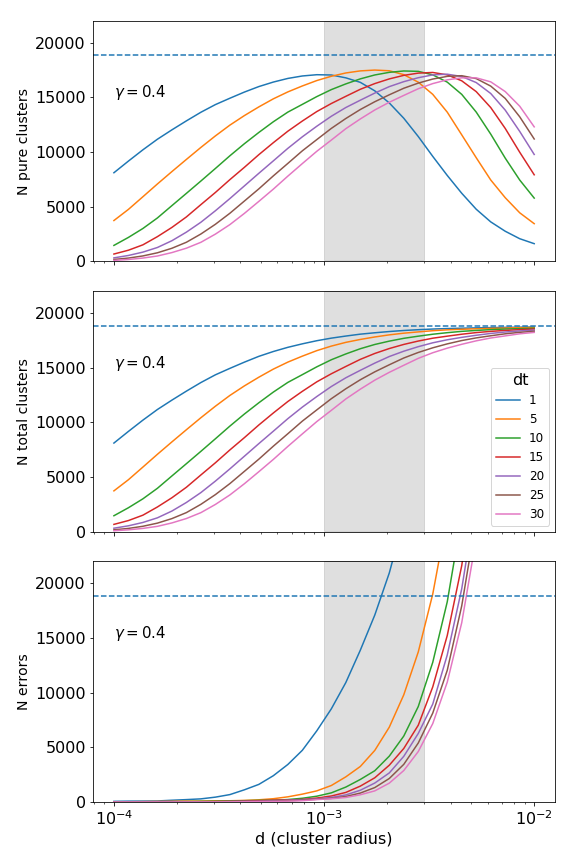}\\
    \caption{%
        The dependence of cluster identification and error rate on the tunable parameters.  The tracklets are taken from $\pm 15$~days of the centers of five non-sequential lunations and include the full sky.  The x-axis is the hyper-parameter, $d$, in radians.
        Line colors label the hyper-parameter, $dt$, (in units of days).
        {\bf Top:} Number of clusters correctly identified (horizontal blue dashed line is the total number of available objects with three or more tracklets in lunation).
        {\bf Middle:} Number of clusters correctly identified including those with at least one erroneous tracklet.
        {\bf Bottom} Number of erroneously identified clusters (fewer than three tracklets from any one object).
        The gray band indicates the range of cluster radii for which the searches are complete but still have a relatively low error rate.
        }
    \label{fig:accuracy1}
\end{figure}

\begin{figure}[htbp]
\centering
    \includegraphics[trim = 0mm 0mm 0mm 0mm, clip, angle=0, width=\columnwidth]{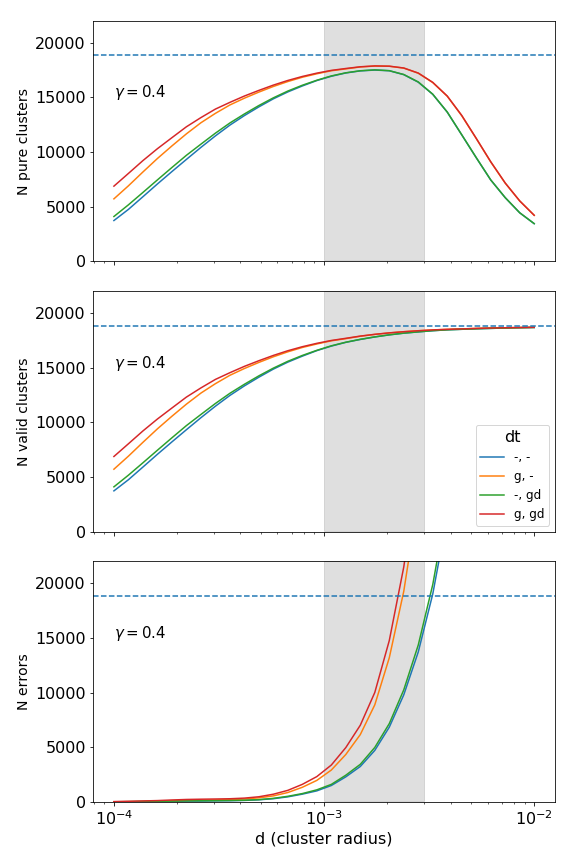}\\
    \caption{%
        The dependence of cluster identification and error rate on $\gamma$ and $\dot \gamma$ spacing, as well as the cluster radius.  The quantities shown in the three panels parallel those in Figure~\ref{fig:accuracy1}.  However, the results are all for $dt=5$.  The four distinct lines show the results for different spacings of $\gamma$ and $\dot\gamma$.  The line indicated as `\_, \_' is the nominal result, for a single distance ($\gamma=0.4$) and five values for the radial velocity ($\dot\gamma=-4\times10^{-3}, -2\times10^{-3}, 0.0, 2\times10^{-3}, 4\times10^{-3}$).  The line indicated as `g, \_' shows the results for increasing the number of $\gamma$ values to $0.3, 0.35, 0.4, 0.45$ but leaving the $\dot\gamma$ values the same as in the nominal run.  The line indicated as `\_, gd' shows the results for using the single $\gamma$ value but making the $\dot\gamma$ values finer ($\dot\gamma=-4\times10^{-3}, -3\times10^{-3}, -2\times10^{-3}, 1\times10^{-3}, 0.0, 1\times10^{-3}, 2\times10^{-3}, 3\times10^{-3}, 4\times10^{-3}$.  And final lined, indicated as `g, gd', increases both the $\gamma$ and $\dot\gamma$ resolution.
        }
    \label{fig:grid_spacing}
\end{figure}

Based on the known identities (labels) of the arrows, we assign one of three different dispositions to each cluster: `pure', `valid', and `erroneous'.  
The arrows in a \emph{pure} cluster all correspond to the same object.  
A \emph{valid} cluster includes three or more arrows from the same object and one or more arrows from another object or objects.  
The underlying object or objects in a valid cluster can in principle be identified through orbit fitting.  
An \emph{erroneous} cluster contains arrows from two or more objects, with no single object having the requisite three arrows.

The top panel of Figure~\ref{fig:accuracy1} shows the number of {\it pure} clusters identified as a function of cluster radius.  The  colored curves show the results for different values of $dt$. The dashed line in each panel shows the total number of objects for which there are at least three tracklets in the data set.  This is the maximum number of objects that can be found.  For small cluster radii, only the tightest clusters are found.  As the cluster radius is increased, more and more clusters are found.  The number of {\it pure} clusters then begins to decrease as larger cluster radii encompass arrows that do not correspond to the same object.  

The middle panel of Figure~\ref{fig:accuracy1} shows the number of {\it valid} clusters.  Again, {\it valid} clusters have at least three arrows of the same object.  In principle, any interloper arrows can be identified and removed via iterative orbit fitting.  {\it Valid} clusters do not become invalid as more interloper arrows are included.  Those clusters simply become more difficult to disentangle.  Thus, the curves in the middle panel asymptote to the maximum number of objects available to be found.

The lower panel of Figure~\ref{fig:accuracy1} shows the number of {\it erroneous} clusters.  As the cluster radius is increased beyond a threshold, the number of erroneous clusters sharply increases.

A range of values of the cluster radius $d$ results in a high degree of completeness with a relatively low error rate.  
The optimum value would depend upon the computational cost required to use orbit fitting to separate interloper tracklets from clusters.  
As a working value, we adopt $d=2\times10^{-3}$~rad.
In Section \ref{s:validation} and Figure \ref{fig:S_curve} we provide further consideration of ways to optimize the choice of $d$ in the face of interloper tracklets.

The value of $dt$ sets the relative importance of two arrows having the same starting location versus their moving parallel to each other.  Smaller values of $dt$ yield a higher degree of completeness at smaller cluster radii.  Larger values of $dt$ reduce the error rate by excluding arrows that are not parallel to the others in a cluster.    The value of $dt$ can also be understood physically.  We found that the ideal ratio corresponds to the time span of the observations.  This matches the positional uncertainty with that from the velocity uncertainty.   This is typically a few days when considering the tracklets from single lunations.   We adopt $dt=5$~days.

In addition to the hyper-parameters $dt$ and $d$, the detection efficiency of the algorithm depends on the spacing in the adopted values of $\gamma$ and $\dot \gamma$.  
The results in Figure~\ref{fig:accuracy1} are from a single value $\gamma=0.4$ and $\Delta \dot \gamma = 2\times10^{-3}$.  Here we explore the effect of using a range of $\gamma$ values and a finer grid in $\dot \gamma$.  
Figure~\ref{fig:grid_spacing} includes the same set of curves as shown in Figure~\ref{fig:accuracy1}, but the different curves show the results for different spacing of the adopted values of $\gamma$ and $\dot \gamma$.  We explored increasing the number of $\gamma$ values from a single value to four values ($\gamma=0.3, 0.35, 0.4, 0.45$) and increasing the $\dot\gamma$ resolution by a factor of two.  What we find is that finer spacing in either parameter improves the completeness but not dramatically so.  We also found that further increasing the resolution results in very little improvement in completeness, as expected.  The tightness of a cluster of arrows is ultimately determined by the observational uncertainties associated with the underlying tracklets and the time elapses between the observations and the reference time.   Once the parameter spacing results in a cluster that is tighter than the observational uncertainties allow, finer parameter spacing is not helpful.

\subsection{Other Orbital Populations}
\label{SECN:POPS} 
In Sections \ref{s:FixedGamma} and \ref{SECN:HYPER1}, we have demonstrated how one can efficiently search for Main Belt Asteroids within the training data-set from the Unnumbered Observations file.

As our clustering method depends on making heliocentric transformations at an asserted inverse-heliocentric distance, $\gamma$, if one wishes to efficiently search for \emph{other} types of solar system object at rather different heliocentric distances (e.g. NEOs, Trojans, Centaurs, TNOs, etc), judicious choices have to be made regarding the parameters used when performing searches. 

In Appendix \ref{a:POPS} we provide a detailed discussion of the optimal parameters (e.g. the assumed $\gamma$, the radius hyper-parameter, $d$, etc) to be used when searching for such objects.


\section{Cluster Validation}
\label{s:validation}

\subsection{Sanity Checks}
\label{s:sanity}
The clusters constructed via the processes described in Section \ref{SECN:DEMO} can produce a large number of small, three-element clusters that exist in close proximity, many of which are proper sub-clusters of other clusters. 
We identify and remove any proper subsets (note that this can be done for any data-set, not just labelled training data).  In addition, we discard any clusters for which all of the arrows come from a single night.  Such clusters are helpful for the purpose of identifying tracklets that correspond to the same object, but they do not help establish an orbit.

The clusters constructed via the processes described in Section \ref{SECN:DEMO} may contain tracklets with duplicate times.
If the duplicate times arise from detections taken in the same exposure at the same observatory, we do {\it not} allow these tracklets to occupy the same cluster. 
We deal with this scenario by identifying the number of tracklets with duplicate times, $N_{T,D}$, and then splitting the cluster into $N_{T,D}$ overlapping subsets, each of which receives {\it one} of the  duplicate-time tracklets and {\it all} of the other (non-overlapping) tracklets.
This process is repeated to exhaustion, ensuring no cluster or sub-cluster contains any duplicate times.

\subsection{Cluster Validation}
\label{s:Valid}
%
%
\begin{figure}[htbp]
\centering
    \includegraphics[trim = 0mm 15mm 0mm 15mm, clip, angle=0, width=\columnwidth]{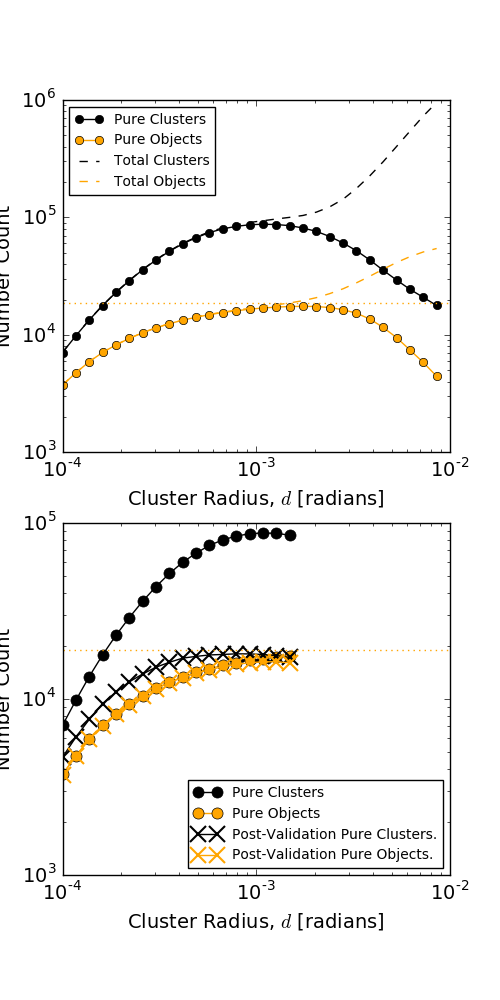}
    \caption{%
        Details of clusters identified for $dt=5$, corresponding to the orange line in Figure \ref{fig:accuracy1}.
        To maintain consistency, data plotted in orange here are unique \emph{objects} found in the training data, while black data are the unique \emph{clusters} constituting those objects.
        {\bf Top: } 
        Dashed lines indicate the \emph{total} identified objects/clusters, while the dots indicate the subset that are \emph{pure}.
        {\bf Bottom: }
        Black and Orange dots are repeated from the top panel.
        Crosses indicate the number of clusters (black) and unique objects (orange) that remained after the simple validation steps of Section \ref{s:validation}. 
        Our validation steps remove many clusters, while leaving at least one cluster associated with almost all pure objects.
        }
    \label{fig:S_curve}
\end{figure}
%

As demonstrated in Section \ref{SECN:DEMO}, some fraction of the clusters generated will be \emph{im}pure, containing tracklets from multiple objects. 
While the selection of appropriate hyper-parameters (e.g., the search radius, $d$) can drastically reduce the generation of impure clusters, we still need to be able to defend against impure clusters. 
A number of different methods can be imagined to identify and reject impure clusters. 
In the remainder of this section we describe one such method.

We emphasize that while one can ultimately perform a full orbit-fit on any cluster generated, at this stage in the analysis we are interested in quickly and cheaply identifying and excluding as many impure clusters as possible. 
Only after that is done do we consider it reasonable to move on to do full orbit-fits.

\subsubsection{6-Dimensional Cluster Refitting}
\label{s:Refit}
The clusters generated in Section \ref{SECN:DEMO} were identified based on their fitted values of $\alpha,\dot\alpha,\beta,\dot\beta$, where those values were generated using \emph{assumed} values of $\gamma$ and $\dot\gamma$.
Having identified the clusters we can now refit each cluster in all 6 parameters, i.e. allowing both $\gamma$ and $\dot\gamma$ to vary as well. 
Allowing $\gamma$ and $\dot\gamma$ to vary means that the two expressions in Equation \ref{rearrange} are no longer independent, hence a non-linear fit is required. 
A number of fitting methods are possible: we find that the {\sc scipy} \citep{SciPy} ``minimize'' function, employing the BFGS algorithm, is sufficient for our requirements. 

As described in Section \ref{SECN:HELIO}, a number of gravity models are possible for evaluating the $g_x(t^\prime), g_y(t^\prime)$ terms in Equation \ref{rearrange}. 
Because we continue to favor efficiency at this stage, we continue to use the $-\frac{1}{2}\sigma t^2$ model used in Section \ref{SECN:DEMO}.

If the 6-dimensional cluster fit is good, the cluster is retained and assumed to be worthy of a more detailed orbit-fit.

\subsubsection{Gauss' Method}
\label{s:Gauss}
If a cluster fails to pass the criteria for refitting described in Section \ref{s:Refit}, we attempt to directly fit the observations using Gauss' Method~\citep{Danby.1992}.  If Gauss' Method then yields an acceptable fit, we retain these additional clusters.

\subsubsection{Results}
\label{s:Results}
Given a set of clusters that have passed either of the fitting methods described in Section \ref{s:Refit} and \ref{s:Gauss}, we could then carry out a full orbit fit. 
Given labelled training data (for which the purity and the orbital characteristics of the objects are already known), this step is superfluous. 

For the training data described in Section \ref{SECN:DEMO} and illustrated in (e.g.) Figure \ref{fig:accuracy1}, we use the results from the $dt=$ curve.  We then perform the steps described in Sections \ref{s:sanity} and \ref{s:Valid}, and plot the results in Figure \ref{fig:S_curve}.  
Figure \ref{fig:S_curve} illustrates three important points:
\begin{enumerate}
    \item \emph{The ``S''-shape of the dashed-black curve. } 
    Further to the detailed training illustrated in Figures \ref{fig:accuracy1} and \ref{fig:grid_spacing} we find that the ``S''-shape of the dashed-black curve can tell us about the appropriate value of $d$ required to conduct a search. 
    In particular, we see that even without using the labels of the data (i.e., using only the black dashed line in the top panel of Figure \ref{fig:S_curve}, and hence without knowing which of the clusters are pure), the changing gradients of the curve indicate regions of different purity.
    Using the yellow, labelled data in the top panel of Figure \ref{fig:S_curve} verifies that towards the left of the curve, the clusters are pure but incomplete, while towards the right of the curve the clusters are complete but highly impure. 
    Hence, a cluster radius around the inflection point of the ``S''-curve  is close to a sweet-spot, where the results are both highly complete and highly pure. 
    This insight will be of value in Section \ref{SECN:ITF} when deciding an appropriate value of $d$ to select when searching for clusters in the ITF data.
    \item \emph{Many Clusters are Removed. }
    In the bottom panel of Figure \ref{fig:S_curve}, the black crosses are significantly below the black dots, indicating that the majority of clusters are removed as a result of the steps described in Sections \ref{s:sanity} and \ref{s:Valid}. 
    This reduces by an order of magnitude the number of clusters that 
    will later require a more expensive full orbit fit. 
    \item \emph{Almost all Objects Remain. }
    Despite the removal of the majority of the \emph{clusters}, many of which were pure, we find that the majority of \emph{objects} remain. 
    I.e., in the bottom panel of Figure \ref{fig:S_curve}, we see that the yellow crosses are almost exactly coincident with the yellow dots. 
    This means that, despite a number of pure clusters being removed, at least one cluster remains for almost all objects, ensuring that we remain highly complete. 
\end{enumerate}

\subsection{Inter-Lunation Linking: Orbit Similarity}
\label{s:INTER}
Using the re-fitted and refined values of \\ $(\alpha, \dot\alpha, \beta, \dot\beta, \gamma, \dot\gamma)$ generated during the clustering-of-clusters, we can transform to Keplerian Elements, $(a,e,i,\Omega,\omega, MA)$.
As is well-known, the first five of these elements are slowly varying, and provide a means to link clusters across lunations (and beyond).
We regard this as essentially a ``solved problem'': once a cluster with three or more tracklets has been established in one lunation, and a reasonable candidate orbit fitted, a variety of efficient methods exist to propagate that orbit to other lunations. 
Hence any associated tracklets and/or clusters in other lunations can be incorporated into the a final fitted orbit solution.


\section{Results: Searching the Isolated Tracklet File}
\label{SECN:ITF}
Having established the performance of our heliocentric clustering algorithm on labelled data (the Unnumbered Observation File) in Sections \ref{SECN:DEMO} and \ref{s:validation}, we now apply our method to search for new objects within the MPC's ``Isolated Tracklet File'' or ITF\footnote{\url{http://www.minorplanetcenter.net/iau/ITF/itf.txt.gz}}.
Roughly 90\% of the tracklets reported to the MPC can be immediately matched with known objects.  
Of the remaining 10\%, many are either linked with other tracklets reported within the previous few days or observed by a follow-up program.  
However, some tracklets are not identified, linked, or immediately re-observed.  
These unmatched tracklets are stored in the ITF, with the hope that they can be linked to future observations.  These are essentially asteroids that have fallen `down the back of the couch.'

\subsection{ITF Data Set}
\label{SECN:ITF:DATA}
At the time of our analysis, the ITF contained about fourteen million observations grouped into nearly four million tracklets.  
Most of those tracklets are real.  
And given the area of sky observed nightly by large NASA-funded surveys, we estimate that most objects have been observed multiple times.  
It is not uncommon for newly discovered objects, with well-determined orbits, to be subsequently found to match several tracklets in the ITF, spread out over a span of several years~(e.g. \citealt{Chen.2016, Weryk17}).

\begin{figure}[htbp]
\centering
    %
    \includegraphics[trim = 0mm 35mm 0mm 30mm, clip, angle=0,     width=\columnwidth]{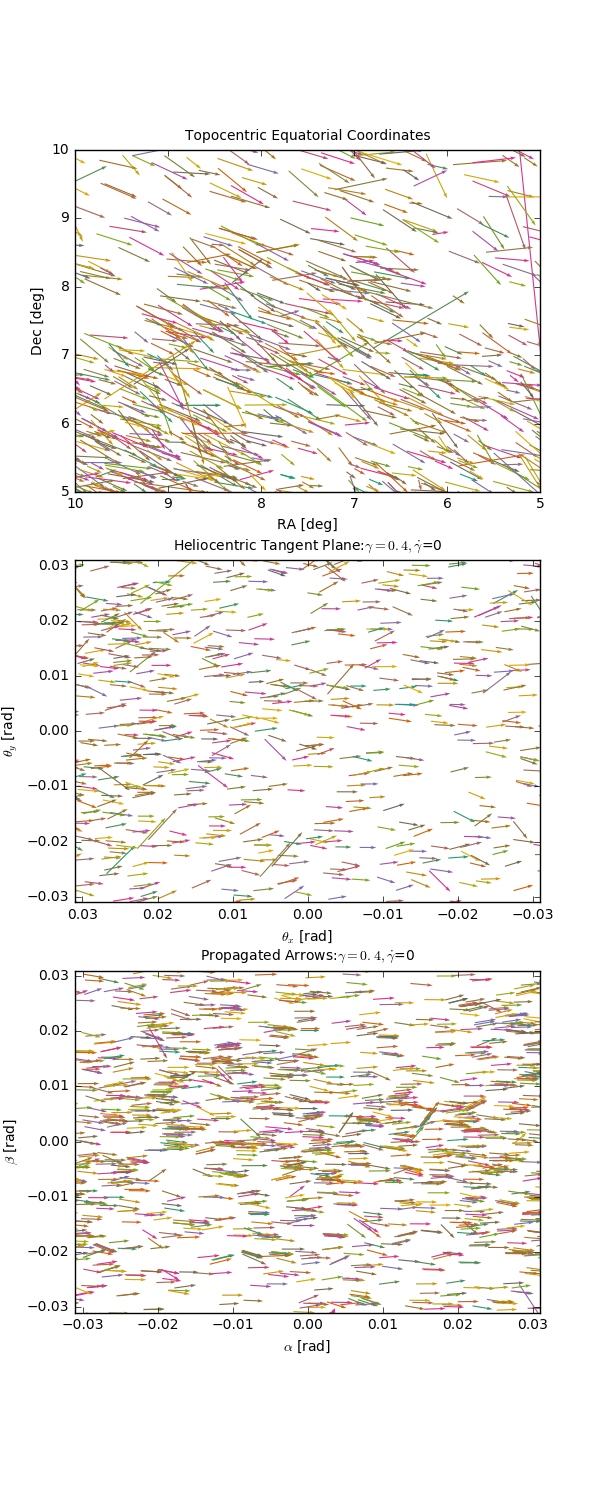}
    \caption{Sample ITF trackles: 30 days of observations in a small portion of the night sky. 
    {\bf Top:} Tracklets plotted in equatorial coordinates;
    {\bf Middle:} Tracklets transformed to heliocentric $\theta_x$ and $\theta_y$ coordinates as per Eqn. \ref{posexpression}, assuming $(\gamma,\dot\gamma)=(0.4,0)$.
    The direction and length of the arrows represent $\dot\theta_x$ and $\dot\theta_y$.
    {\bf Bottom:}
    Using Eqn. \ref{rearrange} we fit for ($\alpha,\beta,\dot\alpha,\dot\beta$).
    The tails of the plotted arrows indicate $\alpha$ and $\beta$, the direction and length represent $\dot\alpha$ and $\dot\beta$.  
    Unlike the labelled data of Fig. \ref{fig:accuracy1}, we cannot color the arrows according to their known identify.  
    Instead we color the arrows according to the cluster that they have been identified with. 
    The propagated arrows in the bottom panel display the same clear ``clustering'' by color as was seen in the labelled data of Figure~\ref{fig:accuracy1}. }
    \label{fig:ITF_sample}
\end{figure}

As an illustration of the ITF data, in Figure \ref{fig:ITF_sample} we provide plots corresponding to those seen in Figure \ref{fig:UnnObs} for the labelled data in Section \ref{SECN:DEMO}.
At the top of Figure \ref{fig:ITF_sample} we plot data in a ``window'' in which the time is within $15$~days of JD 2457308.5.

In the second panel of Figure \ref{fig:ITF_sample}, the tracklets in have been transformed to heliocentric tangent coordinates assuming that $(\gamma,\dot{\gamma}) = ( 0.4, 0.0 ) $.
In the third panel, use Eqn. \ref{rearrange} to fit for ($\alpha,\beta,\dot\alpha,\dot\beta$) and plot their values as arrows, for which we have adopted $dt=5$~days.

Unlike the labelled data of Fig. \ref{fig:UnnObs}, we cannot color the arrows in Figure \ref{fig:ITF_sample} according to their known identify (as this is unknown).  
Instead we color the arrows according to the cluster that they have been identified with. 
After doing this, we see that the propagated arrows in the bottom panel display the same clear ``clustering'' by color as was seen in the labelled data of Fig. \ref{fig:UnnObs}

\subsection{ITF Cluster Radius Determination}
\label{SECN:ITF:S-Curve}
%
\begin{figure}[ht]
\centering
  \includegraphics[trim = 0mm 0mm 0mm 0mm, clip, angle=0, width=\columnwidth]{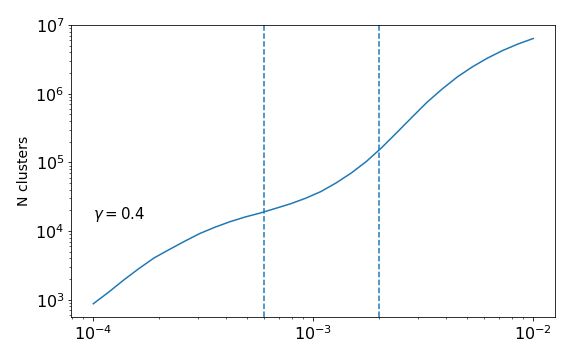}
  \caption{%
    The total number of clusters identified in five ITF lunations, as a function of cluster radius.  The two vertical dashed lines, at $d=6\times10^{-4}$ and $d=2\times10^{-3}$ indicate the cluster radii we used for the searches of the ITF and MPC training data set, respectively.
    }
    \label{fig:ITF:S-curve}
\end{figure}
%
Following the analysis of Section \ref{s:Results} and Figure \ref{fig:S_curve}, we search for clusters in 5 different lunations of the ITF data.
As we did previously, we use a single  value of $\gamma=0.4$ and five values of $\dot\gamma$ evenly spaced from $-4\times10^{-4}$ to $4\times10^{-4}$.
We vary the search radius, $d$, and then plot the total number of returned clusters as a function of $d$ in Figure \ref{fig:ITF:S-curve}.

We find that the curve in Figure \ref{fig:ITF:S-curve} has a similar ``S''-shape to that of the training data in Figure \ref{fig:S_curve}, but we see that the transition to the plateau-region occurs slightly earlier, i.e. at values around $\sim6\times10^{-4}$~rad, rather than the $\sim1\times10^{-3}$~rad seen for the training data. 
We interpret this to mean that the ratio of clusterable-objects (i.e. identifiable objects with at least 3 tracklets per cluster) to background contaminants (i.e. un-clusterable single- and pair-tracklets) is lower in the ITF than it was in the training data.  To keep the number of contaminated clusters to a minimum, while still achieving good completeness, we adopt a smaller cluster radius $d\sim6\times10^{-4}$~rad.

\subsection{Identified Clusters}
\label{SECN:ITF:STATS}
\begin{figure}[ht]
\centering
  \includegraphics[trim = 0mm 0mm 0mm 0mm, clip, angle=0, width=\columnwidth]{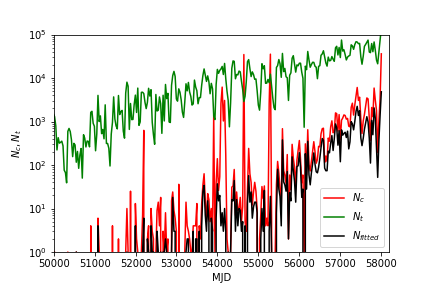}
  \caption{%
    The total number of tracklets (green), the resulting number of clusters prior to vetting (red), and the number of clusters after vetting (black) as a function of lunation for the ITF.  The spikes in the number of clusters prior to vetting are due to large groups of tracklets, in close proximity on the same, from the same observatory on the same nights.  These are apparent observational artifacts. 
    }
    \label{fig:number_clusters_tracklets}
\end{figure}

We now search for clusters in the data using the parameters identified in Section~\ref{SECN:ITF:S-Curve}.  The number of tracklets and resulting clusters (prior to vetting), as a function of lunation, are shown in Figure~\ref{fig:number_clusters_tracklets}.
%

\subsection{Verified Clusters}
\label{SECN:ITF:VERIFY}
We now undertake the validation process described in Section \ref{s:validation}.
This process enables us to reject some of the clusters identified in Section \ref{SECN:ITF:STATS}. 
The number of clusters which survive the validation process is $\sim41,000$.

Following the process described above, we have submitted all validated clusters to the MPC\footnote{Using the process described at\\ \url{https://www.minorplanetcenter.net/iau/ITF/ReadMe.txt}}. 
These clusters will then be independently validated by MPC staff-member Gareth Williams\footnote{As Gareth Williams will act as the MPC's official ``processor'' of these data (acting to verify the validity of the links from the point of view of the MPC and IUA), he wishes to retain a degree of impartiality by {\it not} being named as an author on this paper.}.
Following validation, they will be removed from the ITF.

\section{Discussion}
\label{SECN:DISCUSS}
We have developed a novel algorithm which employs a heliocentric transformation and propagation methods that allow for the ``linking'' of minor-planet tracklets via clustering.

This method scales as $O(N_t \log N_t)$ in the number of tracklets $N_t$, unlike previous methods which typically scale as $O(N_t^3)$ or worse~\citep{Kubica.2007a,Denneau.2013,Jones.2017}.  Thus, it is applicable to data sets with very large numbers of tracklets. In addition, this method is trivially parallelized over different regions of the heliocentric sky, as well over different sets of adopted parameters ($\gamma,\dot{\gamma}$).

This is clearly of great significance for the processing data from upcoming surveys such as LSST, and is likely to enable significant savings in CPU-related costs.
The practicalities of processing LSST data using our clustering approach is left to future work.

Going beyond the specific implementation demonstrated here, our method can be applied to more widely separated tracklets, those in separate lunations, by including a more accurate gravity model.   In addition, we note that our approach can be generalized to searches for single detections by searching over values for $\dot\alpha$ and $\dot\beta$ in addition to $\gamma$ and $\dot\gamma$.  This would be analogous to image-stacking searches for faint moving objects that scan over rate of motion or orbital parameters~\citep{Gladman.2001,Holman.2004,Bernstein.2004}.  Such an approach would preserve the efficient scaling presented here. 

\acknowledgments

MJH and MJP gratefully acknowledge 
NASA grants NNX12AE89G, NNX16AD69G, and \newline 
NNX17AG87G, as well as support from the Smithsonian 2015-2017 Scholarly Studies program.
We are grateful to Brian Plancher and the other Harvard CS182 teaching staff members for their support and guidance.  
We thank Gareth Williams of the MPC for orbit fitting tools and for his invaluable help in validating our clustered objects.
We have also benefited from helpful discussions with Timothy Spahr, Jonathan Myers, and David Gerdes.


\begin{thebibliography}{}
\expandafter\ifx\csname natexlab\endcsname\relax\def\natexlab#1{#1}\fi
\providecommand{\url}[1]{\href{#1}{#1}}

\bibitem[{{Bannister} {et~al.}(2016){Bannister}, {Kavelaars}, {Petit},
  {Gladman}, {Gwyn}, {Chen}, {Volk}, {Alexandersen}, {Benecchi}, {Delsanti},
  {Fraser}, {Granvik}, {Grundy}, {Guilbert-Lepoutre}, {Hestroffer}, {Ip},
  {Jakubik}, {Jones}, {Kaib}, {Kavelaars}, {Lacerda}, {Lawler}, {Lehner},
  {Lin}, {Lister}, {Lykawka}, {Monty}, {Marsset}, {Murray-Clay}, {Noll},
  {Parker}, {Pike}, {Rousselot}, {Rusk}, {Schwamb}, {Shankman}, {Sicardy},
  {Vernazza}, \& {Wang}}]{Bannister.2016}
{Bannister}, M.~T., {Kavelaars}, J.~J., {Petit}, J.-M., {et~al.} 2016, \aj,
  152, 70

\bibitem[{{Bernstein} \& {Khushalani}(2000)}]{Bernstein.2000}
{Bernstein}, G., \& {Khushalani}, B. 2000, \aj, 120, 3323

\bibitem[{{Bernstein} {et~al.}(2004){Bernstein}, {Trilling}, {Allen}, {Brown},
  {Holman}, \& {Malhotra}}]{Bernstein.2004}
{Bernstein}, G.~M., {Trilling}, D.~E., {Allen}, R.~L., {et~al.} 2004, \aj, 128,
  1364

\bibitem[{{Chen} {et~al.}(2016){Chen}, {Lin}, {Holman}, {Payne}, {Fraser},
  {Lacerda}, {Ip}, {Chen}, {Kudritzki}, {Jedicke}, {Wainscoat}, {Tonry},
  {Magnier}, {Waters}, {Kaiser}, {Wang}, \& {Lehner}}]{Chen.2016}
{Chen}, Y.-T., {Lin}, H.~W., {Holman}, M.~J., {et~al.} 2016, \apjl, 827, L24

\bibitem[{{Christensen} {et~al.}(2016){Christensen}, {Carson Fuls}, {Gibbs},
  {Grauer}, {Johnson}, {Kowalski}, {Larson}, {Leonard}, {Matheny}, {Seaman}, \&
  {Shelly}}]{Christensen.2016}
{Christensen}, E.~J., {Carson Fuls}, D., {Gibbs}, A., {et~al.} 2016, in
  AAS/Division for Planetary Sciences Meeting Abstracts, Vol.~48, AAS/Division
  for Planetary Sciences Meeting Abstracts, 405.01

\bibitem[{Curtin {et~al.}(2013)Curtin, March, Ram, Anderson, Gray, \&
  Jr.}]{Curtin.2017}
Curtin, R.~R., March, W.~B., Ram, P., {et~al.} 2013, CoRR, abs/1304.4327,
  arXiv:1304.4327.
\newblock \url{http://arxiv.org/abs/1304.4327}

\bibitem[{{Danby}(1992)}]{Danby.1992}
{Danby}, J.~M.~A. 1992, {Fundamentals of celestial mechanics}

\bibitem[{{Denneau} {et~al.}(2013){Denneau}, {Jedicke}, {Grav}, {Granvik},
  {Kubica}, {Milani}, {Vere{\v s}}, {Wainscoat}, {Chang}, {Pierfederici},
  {Kaiser}, {Chambers}, {Heasley}, {Magnier}, {Price}, {Myers}, {Kleyna},
  {Hsieh}, {Farnocchia}, {Waters}, {Sweeney}, {Green}, {Bolin}, {Burgett},
  {Morgan}, {Tonry}, {Hodapp}, {Chastel}, {Chesley}, {Fitzsimmons}, {Holman},
  {Spahr}, {Tholen}, {Williams}, {Abe}, {Armstrong}, {Bressi}, {Holmes},
  {Lister}, {McMillan}, {Micheli}, {Ryan}, {Ryan}, \& {Scotti}}]{Denneau.2013}
{Denneau}, L., {Jedicke}, R., {Grav}, T., {et~al.} 2013, \pasp, 125, 357

\bibitem[{{Gerdes} {et~al.}(2017){Gerdes}, {Sako}, {Hamilton}, {Zhang},
  {Khain}, {Becker}, {Annis}, {Wester}, {Bernstein}, {Scheibner}, {Zullo},
  {Adams}, {Bergin}, {Walker}, {Mueller}, {Abbott}, {Abdalla}, {Allam},
  {Bechtol}, {Benoit-L{\'e}vy}, {Bertin}, {Brooks}, {Burke}, {Carnero Rosell},
  {Carrasco Kind}, {Carretero}, {Cunha}, {da Costa}, {Desai}, {Diehl},
  {Eifler}, {Flaugher}, {Frieman}, {Garc{\'{\i}}a-Bellido}, {Gaztanaga},
  {Goldstein}, {Gruen}, {Gschwend}, {Gutierrez}, {Honscheid}, {James}, {Kent},
  {Krause}, {Kuehn}, {Kuropatkin}, {Lahav}, {Li}, {Maia}, {March}, {Marshall},
  {Martini}, {Menanteau}, {Miquel}, {Nichol}, {Plazas}, {Romer}, {Roodman},
  {Sanchez}, {Sevilla-Noarbe}, {Smith}, {Smith}, {Soares-Santos}, {Sobreira},
  {Suchyta}, {Swanson}, {Tarle}, {Tucker}, {Zhang}, \& {DES
  Collaboration}}]{Gerdes.2017}
{Gerdes}, D.~W., {Sako}, M., {Hamilton}, S., {et~al.} 2017, \apjl, 839, L15

\bibitem[{{Gladman} {et~al.}(2001){Gladman}, {Kavelaars}, {Petit},
  {Morbidelli}, {Holman}, \& {Loredo}}]{Gladman.2001}
{Gladman}, B., {Kavelaars}, J.~J., {Petit}, J.-M., {et~al.} 2001, \aj, 122,
  1051

\bibitem[{{G{\'o}rski} {et~al.}(2005){G{\'o}rski}, {Hivon}, {Banday},
  {Wandelt}, {Hansen}, {Reinecke}, \& {Bartelmann}}]{Gorski.2005}
{G{\'o}rski}, K.~M., {Hivon}, E., {Banday}, A.~J., {et~al.} 2005, \apj, 622,
  759

\bibitem[{{Holman} {et~al.}(2004){Holman}, {Kavelaars}, {Grav}, {Gladman},
  {Fraser}, {Milisavljevic}, {Nicholson}, {Burns}, {Carruba}, {Petit},
  {Rousselot}, {Mousis}, {Marsden}, \& {Jacobson}}]{Holman.2004}
{Holman}, M.~J., {Kavelaars}, J.~J., {Grav}, T., {et~al.} 2004, \nat, 430, 865

\bibitem[{{Holman} {et~al.}(2017){Holman}, {Payne}, {Fraser}, {Lacerda},
  {Bannister}, {Lackner}, {Chen}, {Lin}, {Smith}, {Kotanekova}, {Young},
  {Chambers}, {Chastel}, {Denneau}, {Fitzsimmons}, {Flewelling}, {Grav},
  {Huber}, {Induni}, {Kudritzki}, {Krolewski}, {Jedicke}, {Kaiser}, {Lilly},
  {Magnier}, {Mark}, {Meech}, {micheli}, {Murray}, {Parker}, {Protopapas},
  {Ragozzine}, {Veres}, {Wainscoat}, {Waters}, \& {Weryk}}]{Holman.2017}
{Holman}, M.~J., {Payne}, M.~J., {Fraser}, W., {et~al.} 2017, ArXiv e-prints,
  arXiv:1709.05427

\bibitem[{Jones {et~al.}(2001)Jones, Oliphant, Peterson, {et~al.}}]{SciPy}
Jones, E., Oliphant, T., Peterson, P., {et~al.} 2001, {SciPy}: Open source
  scientific tools for {Python}, , , [Online; accessed <today>].
\newblock \url{http://www.scipy.org/}

\bibitem[{{Jones} {et~al.}(2017){Jones}, {Slater}, {Moeyens}, {Allen},
  {Axelrod}, {Cook}, {Ivezi{\'c}}, {Juri{\'c}}, {Myers}, \&
  {Petry}}]{Jones.2017}
{Jones}, R.~L., {Slater}, C.~T., {Moeyens}, J., {et~al.} 2017, ArXiv e-prints,
  arXiv:1711.10621

\bibitem[{{Kubica} {et~al.}(2007){Kubica}, {Denneau}, {Grav}, {Heasley},
  {Jedicke}, {Masiero}, {Milani}, {Moore}, {Tholen}, \&
  {Wainscoat}}]{Kubica.2007a}
{Kubica}, J., {Denneau}, L., {Grav}, T., {et~al.} 2007, \icarus, 189, 151

\bibitem[{{Kulkarni}(2016)}]{Kulkarni.2016}
{Kulkarni}, S.~R. 2016, {The Zwicky Transient Facility}, ,

\bibitem[{{Lin} {et~al.}(2016){Lin}, {Chen}, {Holman}, {Ip}, {Payne},
  {Lacerda}, {Fraser}, {Gerdes}, {Bieryla}, {Sie}, {Chen}, {Burgett},
  {Denneau}, {Jedicke}, {Kaiser}, {Magnier}, {Tonry}, {Wainscoat}, \&
  {Waters}}]{Lin.2016}
{Lin}, H.~W., {Chen}, Y.-T., {Holman}, M.~J., {et~al.} 2016, \aj, 152, 147

\bibitem[{{Mainzer} {et~al.}(2011){Mainzer}, {Grav}, {Bauer}, {Masiero},
  {McMillan}, {Cutri}, {Walker}, {Wright}, {Eisenhardt}, {Tholen}, {Spahr},
  {Jedicke}, {Denneau}, {DeBaun}, {Elsbury}, {Gautier}, {Gomillion}, {Hand},
  {Mo}, {Watkins}, {Wilkins}, {Bryngelson}, {Del Pino Molina}, {Desai},
  {G{\'o}mez Camus}, {Hidalgo}, {Konstantopoulos}, {Larsen}, {Maleszewski},
  {Malkan}, {Mauduit}, {Mullan}, {Olszewski}, {Pforr}, {Saro}, {Scotti}, \&
  {Wasserman}}]{Mainzer.2011}
{Mainzer}, A., {Grav}, T., {Bauer}, J., {et~al.} 2011, \apj, 743, 156

\bibitem[{{Mainzer} \& {NEOCam Science Team}(2017)}]{Mainzer.2017}
{Mainzer}, A.~K., \& {NEOCam Science Team}. 2017, in AAS/Division for Planetary
  Sciences Meeting Abstracts, Vol.~49, AAS/Division for Planetary Sciences
  Meeting Abstracts, 219.01

\bibitem[{{Meech} {et~al.}(2017){Meech}, {Weryk}, {Micheli}, {Kleyna},
  {Hainaut}, {Jedicke}, {Wainscoat}, {Chambers}, {Keane}, {Petric}, {Denneau},
  {Magnier}, {Berger}, {Huber}, {Flewelling}, {Waters}, {Schunova-Lilly}, \&
  {Chastel}}]{Meech.2017}
{Meech}, K.~J., {Weryk}, R., {Micheli}, M., {et~al.} 2017, \nat, 552, 378

\bibitem[{{Milani}(1999)}]{Milani.1999}
{Milani}, A. 1999, \icarus, 137, 269

\bibitem[{{Perdelwitz} {et~al.}(2018){Perdelwitz}, {V{\"o}lschow}, \&
  {M{\"u}ller}}]{Perdelwitz.2018}
{Perdelwitz}, V., {V{\"o}lschow}, M., \& {M{\"u}ller}, H.~M. 2018, ArXiv
  e-prints, arXiv:1805.01203

\bibitem[{{Sheppard} \& {Trujillo}(2016)}]{Sheppard.2016}
{Sheppard}, S.~S., \& {Trujillo}, C. 2016, \aj, 152, 221

\bibitem[{{Szalay} {et~al.}(2007){Szalay}, {Gray}, {Fekete}, {Kunszt}, {Kukol},
  \& {Thakar}}]{Szalay.2007}
{Szalay}, A.~S., {Gray}, J., {Fekete}, G., {et~al.} 2007, eprint
  arXiv:cs/0701164, cs/0701164

\bibitem[{{Trujillo} \& {Sheppard}(2014)}]{Trujillo.2014}
{Trujillo}, C.~A., \& {Sheppard}, S.~S. 2014, \nat, 507, 471

\bibitem[{{Vere{\v s}} \& {Chesley}(2017{\natexlab{a}})}]{Veres.2017a}
{Vere{\v s}}, P., \& {Chesley}, S.~R. 2017{\natexlab{a}}, \aj, 154, 12

\bibitem[{{Vere{\v s}} \& {Chesley}(2017{\natexlab{b}})}]{Veres.2017b}
---. 2017{\natexlab{b}}, \aj, 154, 13

\bibitem[{{Weryk} {et~al.}(2017){Weryk}, {Wainscoat}, \& {Williams}}]{Weryk17}
{Weryk}, R.~J., {Wainscoat}, R.~J., \& {Williams}, G. 2017, in AAS/Division for
  Planetary Sciences Meeting Abstracts, Vol.~49, AAS/Division for Planetary
  Sciences Meeting Abstracts \#49, 103.02

\end{thebibliography}


\appendix

\section{Heliocentric Transformation}
\label{APP:HELIO}
Let us first assume a value for $r$, the heliocentric distance to the object.  This relates the topocentric distance to the object $\rho$, the heliocentric position of the observatory $r_{obs}$, and the solar elongation $\phi$ as follow:
\begin{equation}
r^2 = \rho^2 + {r_{obs}}^2 - 2 \rho r_{obs} \cos \phi
\label{EQN:R}
\end{equation}
where $\cos\phi= - \hat{\rho} \cdot \hat{r}_{obs}$.  After rearranging we get
\begin{equation}
\rho^2 - 2 \rho r_{obs} \cos \phi + {r_{obs}}^2 - r^2 = 0.
\label{EQN:RHOSQ}
\end{equation}
The equation permits zero, one, or two real solutions for $\rho$.
\begin{eqnarray}
\rho &=& r_{obs}\cos \phi \pm \sqrt{{r_{obs}^2 \cos^2 \phi - {r_{obs}}^2 + r^2}}\nonumber\\
     &=& r_{obs}\cos \phi \pm \sqrt{r^2 - {r_{obs}}^2 \sin^2 \phi}\label{EQN:RHO}
\end{eqnarray}
We ignore solutions for which $\rho < 0 $, which implies the observer is looking in the opposite direction (i.e. through the Earth).  
The assumed geometry in these transformations is illustrated in Figure \ref{FIG:CARTOON}, and which amusingly also demonstrates why Eqn.~\ref{EQN:RHO} has the same basic form as the piston motion equations \citep[e.g.][]{weston1992energy}.

Given a solution for $\rho$, the heliocentric position of the target is 
\begin{equation}
    {\bf r} = {\bf r_{obs}} + \rho {\bf \hat{\rho}}.
\end{equation}

\section{Other Orbital Populations}
\label{a:POPS} 
The training in Section \ref{SECN:HYPER1} used a heterogeneous data set, with tracklets from a variety of minor-planet populations.
We now wish to understand the efficiency of identification of a range of different population classes, and in particular, to identify the appropriate values of $\gamma,\dot\gamma,d$ and $dt$ that allow for the most efficient recovery of each population.
We provide results for NEO, Trojan, Centaur and TNO populations (the data in Figures \ref{fig:UnnObs} and \ref{fig:focus} is dominated by MBAs, hence Section \ref{SECN:HYPER1} and Figure \ref{fig:accuracy1} suffice to characterize the linking of MBAs).

For each of the NEO, Trojan, Centaur, and TNO populations we create training sets composed purely of each type of object.
The small size and sparse nature of this NEO training set means that our error rates will be artificially low because there are far fewer tracklets that might be transformed in a way that contaminates another cluster.
Nevertheless, we select appropriate ranges of $\gamma$ and $\dot\gamma$ for each population (see below), and investigate the recovery of the labelled objects under various hyper-parameter choices.

\subsection{NEOs}
\label{a:NEOs}
%
\begin{figure}[htbp]
\centering
    \includegraphics[trim = 0mm 0mm 0mm 0mm, clip, angle=0, width=0.4\columnwidth]{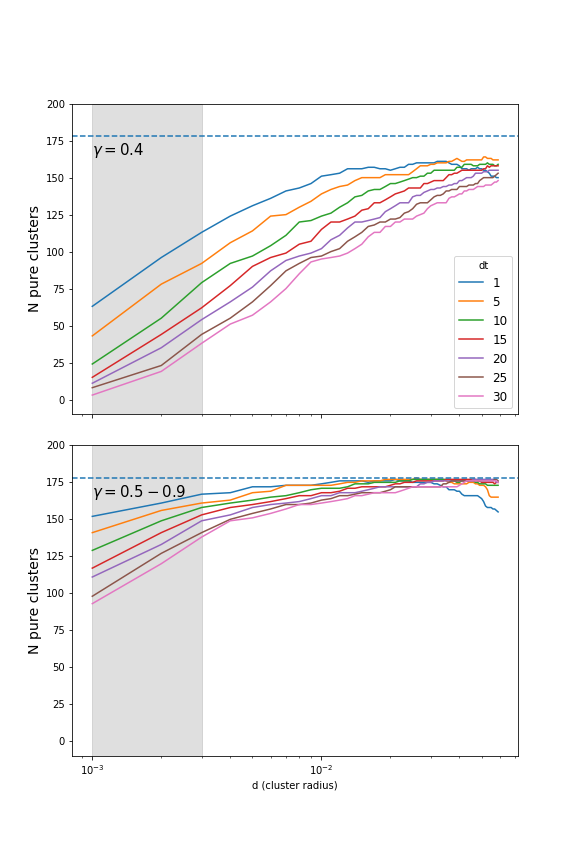}
    \caption{%
        Recovery of NEO tracklets. 
        Line colors label the hyper-parameter, $dt$, (in units of days).
        {\bf Top:} NEO recovery using $\gamma=0.4$, i.e. more suitable for MBAs: gray swath indicates parameters expected for a standard MBA search (see Figure \ref{fig:accuracy1}).
        {\bf Bottom} NEO recovery using $\gamma=0.5-0.9$: gray swath indicates parameters optimized for NEO recover. 
        We find that our method is both remarkably robust and remarkably thorough.
        Even using an ``incorrect'' $\gamma=0.4$, we recover a significant fraction of the NEOs, meaning that a standard ``sweep'' for MBAs would also find half of all NEOs as a fortuitous side-effect.  A larger cluster search radius, even with $\gamma=0.4$, would identify most NEOs.
        To recover the remainder, we would use transformations at higher $\gamma$: $0.5-0.9$ and use a broader search radius.
        }
    \label{fig:NEOaccuracy}
\end{figure}
%
We begin by searching for NEOs using the standard parameters established for MBAs in Section \ref{SECN:HYPER1}.
As illustrated in the top panel of Figure \ref{fig:NEOaccuracy}, we find that our standard method is both remarkably robust and thorough.
Even using an ``incorrect'' $\gamma=0.4$, we recover a significant fraction of the NEOs, meaning that a standard ``sweep'' for MBAs would also find nearly half of all NEOs as a fortuitous side-effect.

Because NEOs can have a large range semi-major axes and may not be near the Earth at the time of discovery, they occupy a large range of parameter space for our $\gamma$ parameter.  A thorough search for NEOs requires that we examine a wider range of $\gamma$ and $\dot\gamma$. We use $0.5\leq\gamma\leq0.9$ in increments of 0.1).  
We show in the bottom panel of Figure \ref{fig:NEOaccuracy} that a larger cluster radius is necessary to recover as many NEOs as possible.

\subsection{Trojans, Centaurs and TNOs}
\label{a:TCT}
%
\begin{figure*}[htbp]
\begin{minipage}[b]{\textwidth}
\centering
    \includegraphics[trim = 0mm 22mm 0mm 0mm, clip, angle=0, width=0.33\textwidth]{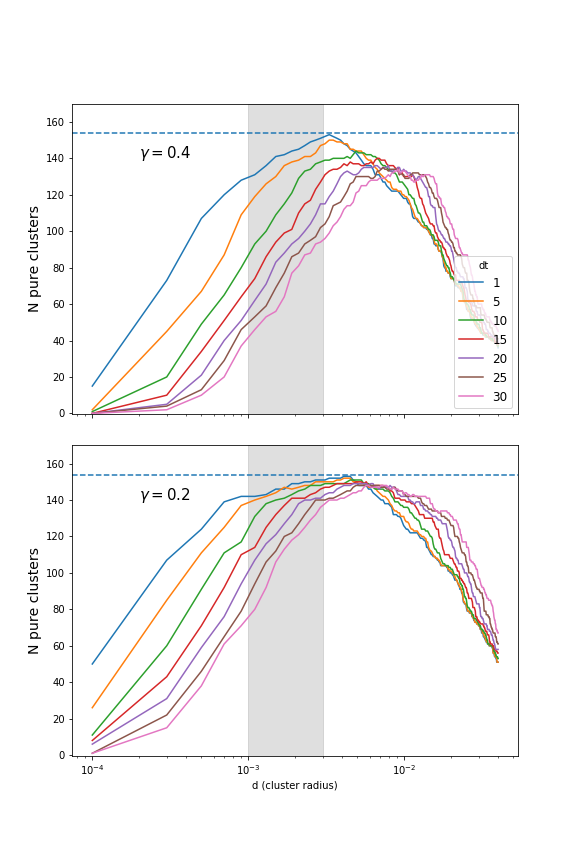}
    \includegraphics[trim = 0mm 22mm 0mm 0mm, clip, angle=0, width=0.33\textwidth]{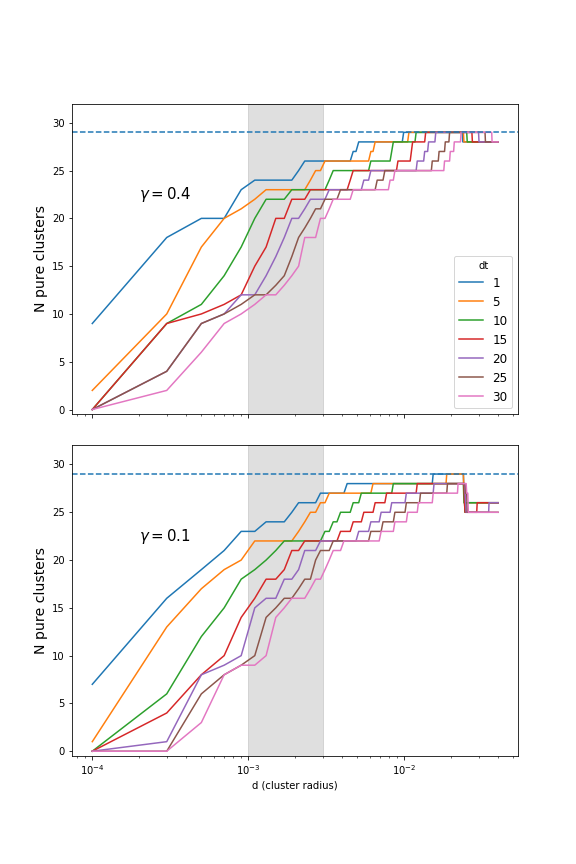}
    \includegraphics[trim = 0mm 22mm 0mm 0mm, clip, angle=0, width=0.33\textwidth]{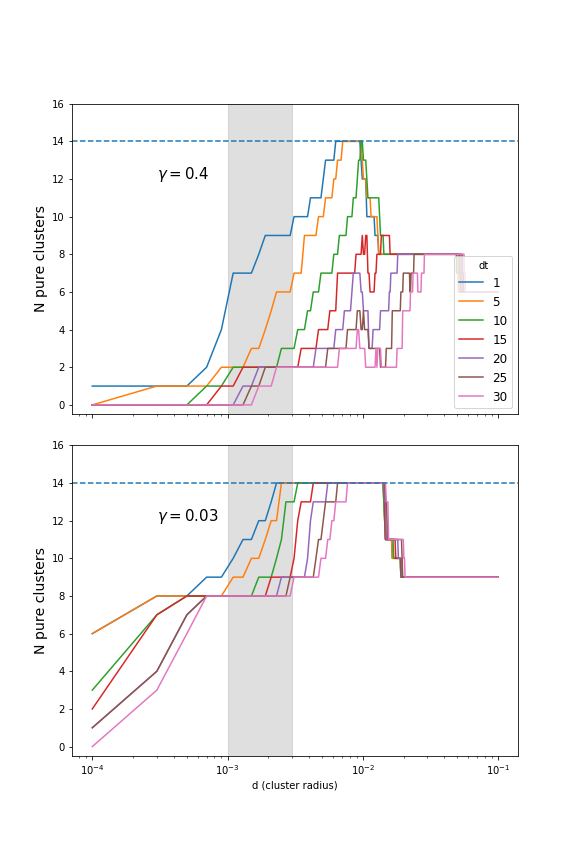}
    \caption{%
        Recovery of Trojans (left), Centaurs (middle) and TNO (right) tracklets. 
        Line colors label the hyper-parameter, $dt$, (in units of days).
        {\bf Top Row:} Recovery using $\gamma=0.4$, i.e. more suitable for MBAs: gray swath indicates parameters expected for a standard MBA search (see Figure \ref{fig:accuracy1}).
        {\bf Bottom Row:} Recovery using appropriate $\gamma$ factors for each population: gray swath indicates parameters optimized for population recovery. 
        As seen for NEOs, the majority of Trojans, Centaurs and TNOs can be recovered during a standard ($\gamma=0.4$) MBA search.  However, these searches are more complete when values of $\gamma$ and ranges of $\dot\gamma$ that are tailored to each distance classes are used.
        }
    \label{fig:MULTIaccuracy}
\end{minipage}
\end{figure*}
%
For each of the Trojan, Centaur and TNO population samples, we begin by establishing the fraction of recovered objects when we search using the standard parameters established for MBAs in Section \ref{SECN:HYPER1}.
We  find that a significant fraction of all objects in each population will be recovered by a standard ``MBA search'' (gray swath, top panels, Figure \ref{fig:MULTIaccuracy}).
Using custom values of $\gamma$, and correspondingly $\dot\gamma$, for each population, we demonstrate in the bottom panels of Figure \ref{fig:MULTIaccuracy} that a more complete recovery of objects in the population can be achieved.

\subsection{Heliocentric Rates of Motion}
\label{a:RatesOfMotion}
%
\begin{figure}[htbp]
\begin{minipage}[b]{\columnwidth}
\centering
    \includegraphics[trim = 0mm 0mm 0mm 0mm, clip, angle=0, width=\columnwidth]{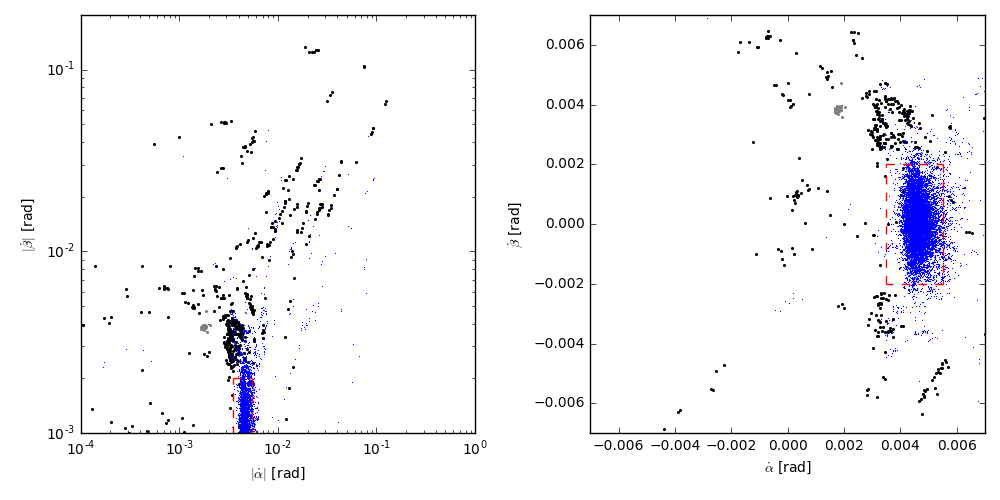}
    \caption{%
        Heliocentric rates of motion, $\dot\alpha$ and $\dot\beta$ for NEOs (black), MBAs (blue) and TNOs (gray) for a transformation using $(\gamma,\dot\gamma)=(0.4,0)$.
        The majority of the MBAs clearly cluster in a region $3.5\times10^{-3}<\dot\alpha<5.5\times10^{-3}$ and $|\dot\beta|<2\times10^{-3}$ (red).
        Excluding objects in this region from {\it subsequent} transformation runs has the effect of significantly reducing both the error rate and the computational load, allowing us to more efficiently find specific, less frequent dynamical classes of objects( NEOs, Trojans, Centaurs, and TNOs).
        }
    \label{fig:rates}
\end{minipage}
\end{figure}
%

The population-specific investigations in Sections \ref{a:NEOs} and \ref{a:TCT} above had no ``background'' population of either main-belt asteroids or false tracklets. 
If such a background were present, the large cluster radii needed to identify some specific populations could drive exceedingly high rates of erroneous clusters, as can be seen from Figure~\ref{fig:accuracy1}.
In a realistic search, we need a means to first remove the large population of main-belt asteroids.  There are a number of possible approaches.  One could first search with parameters appropriate for MBAs, extract the tracklets for the objects linked using those parameters, and the search the smaller set of remaining tracklets with other sets of parameters.
We describe such a detailed fitting and verification procedure  in Section \ref{s:validation} below, but we prefer a rapid, but more approximate, method that allows us to process the entire data set quickly and efficiently.

Different dynamical classes of objects are characterized by their rates of motion.  We can used the fitted heliocentric rates of motion, $\dot\alpha$ and $\dot\beta$, derived in the initial preliminary transformation assuming $\gamma=0.4,\dot\gamma=0$, to identify those tracklets/arrows that likely correspond to MBAs.  

In Figure \ref{fig:rates}, we plot $\dot\alpha$ and $\dot\beta$, corresponding to the arrows already plotted in Figure \ref{fig:UnnObs}. 
We now color the points according to the {\it type} of object (NEO, MBA, etc).
One can clearly see that the majority of MBAs have $\dot\alpha$ and $\dot\beta$ values confined to a region $3.5\times10^{-3}<\dot\alpha<5.5\times10^{-3}$ and $|\dot\beta|<2\times10^{-3}$. 

We can then exclude the tracklets that fall in the MBA region from any subsequent searches at different values of $\gamma$, i.e. when we are searching for different {\it types} of object.   We note that these rates do not strongly depend on solar elongation because the coordinates are heliocentric, assuming the chosen value of $\gamma$ is approximately correct.


\end{document}